\documentclass[
aps,
nofootinbib,
prd,
superscriptaddress,
tightenlines,
notitlepage,
twocolumn,
showpacs,
floatfix
]{revtex4-1}

\usepackage[colorlinks]{hyperref}
\usepackage{tabularx}
\usepackage{graphicx}
\usepackage{amssymb,amsmath,bm,tensor,braket,mathtools}
\usepackage{xcolor}
\usepackage[varg]{txfonts}
\usepackage{enumerate}
\usepackage{mathtools}
\usepackage[capitalize]{cleveref}
\usepackage[utf8]{inputenc}
\usepackage[normalem]{ulem}

\def\al{\alpha}
\def\be{\beta}
\def\ga{\gamma}
\def\de{\delta}
\def\ep{\epsilon}

\def\th{\theta}

\def\ka{\kappa}
\def\la{\lambda}

\def\rh{\rho}

\def\si{\sigma}

\def\ph{\phi}

\def\om{\omega}
\def\Ga{\Gamma}
\def\De{\Delta}

\def\Ph{\Phi}

\def\Om{\Omega}

\def\mn{{\mu\nu}}
\def\abgd{{\al\be\ga\de}}

\def\lsim{\mathrel{\rlap{\lower4pt\hbox{\hskip1pt$\sim$}}
    \raise1pt\hbox{$<$}}}
\def\gsim{\mathrel{\rlap{\lower4pt\hbox{\hskip1pt$\sim$}}
    \raise1pt\hbox{$>$}}}
\def\sqr#1#2{{\vcenter{\vbox{\hrule height.#2pt
         \hbox{\vrule width.#2pt height#1pt \kern#1pt
         \vrule width.#2pt}
         \hrule height.#2pt}}}}

\def\prt{\partial}
\def\lrpartial{\raise 1pt\hbox{$\stackrel\leftrightarrow\partial$}}

\def\part2{\partial_\alpha \partial^\alpha}

\def\pt#1{\phantom{#1}}

\def\xx'{|\vec x -\vec x'|}

\def\b2{b^\al b_\al}

\newcommand{\beq}{\begin{equation}}
\newcommand{\eeq}{\end{equation}}
\newcommand{\bea}{\begin{eqnarray}}
\newcommand{\eea}{\end{eqnarray}}
\newcommand{\bit}{\begin{itemize}}
\newcommand{\eit}{\end{itemize}}
\newcommand{\rf}[1]{(\ref{#1})}
\newcommand\bw{\begin{widetext}}
\newcommand\ew{\end{widetext}}

\newcommand{\DOA}{\affiliation{Department of Astronomy, School of Physics,
Peking University, Beijing 100871, China} }
\newcommand{\KIAA}{\affiliation{Kavli Institute for Astronomy and
Astrophysics, Peking University, Beijing 100871, China}}

\newcommand{\NAOC}{\affiliation{National Astronomical Observatories,
Chinese Academy of Sciences, Beijing 100012, China}}

\begin{document}

\title{Neutron stars in massive scalar-Gauss-Bonnet gravity: Spherical structure
\\ and time-independent perturbations}

\date{\today}
\author{Rui Xu}\email[Corresponding author: ]{xuru@pku.edu.cn}\KIAA
\author{Yong Gao}\DOA\KIAA
\author{Lijing Shao}\email[Corresponding author: ]{lshao@pku.edu.cn}\KIAA\NAOC

\begin{abstract} 

The class of scalar-tensor theories with the scalar field coupling to the Gauss-Bonnet invariant has drawn great interest since solutions of spontaneous scalarization were found for black holes in these theories. We contribute to the existing literature a detailed study of the spontaneously scalarized neutron stars (NSs) in a typical theory where the coupling function of the scalar field takes the quadratic form and the scalar field is massive. The investigation here includes the spherical solutions of the NSs as well as their perturbative properties, namely the tidal deformability and the moment of inertia, treated in a unified and extendable way under the framework of spherical decomposition.
We find that while the mass, the radius, and the moment of inertia of the spontaneously scalarized NSs show very moderate deviations from those of the NSs in general relativity (GR), the tidal deformability exhibits significant differences between the solutions in GR and the solutions of spontaneous scalarization for certain values of the parameters in the scalar-Gauss-Bonnet theory. As a result, the celebrated universal relation between the moment of inertia and the tidal deformability of neutron stars breaks down.
With the mass and the tidal deformability of NSs attainable in the gravitational waves from binary NS mergers, the radius measurable using the X-ray satellites, and the moment of inertia accessible via the high-precision pulsar timing techniques, future multi-messenger observations can be contrasted with the theoretical results and provide us necessary information for building up theories beyond GR.

\end{abstract}

\maketitle
\allowdisplaybreaks  
 
\section{Introduction}
\label{sec:intro}
One aspect of the elegance in Einstein's general relativity (GR) is that the geometric part of the field equations comes from the Einstein-Hilbert action, which merely consists of the simplest curvature invariant of the spacetime manifold, namely, the Ricci scalar. This simplicity is sacrificed in many alternative theories of gravity, where other curvature invariants are taken in to give modifications to GR, attempting to resolve its fundamental incompatibility with quantum field theory \cite{Stelle:1976gc}, or to address issues in cosmology including the initial singularity of the Big Bang and the mystery of dark energy \cite{Starobinsky:1980te, Copeland:2006wr, Nojiri:2010wj}. Among the quadratic curvature invariants, the Gauss-Bonnet term
\bea
{\mathcal G} = -G_\abgd R^\abgd = R_{\abgd}R^{\abgd} - 4 R_{\al\be} R^{\al\be} + R^2 ,
\eea
is special in the sense that adding it multiplied by a constant to the action does not alter the field equations. In the above expression, the double dual of the Riemann curvature tensor is defined as
\bea
G_\abgd := \frac{1}{4} E_{\al\be\ka\la} E_{\ga\de\rh\si} R^{\ka\la\rh\si},
\label{Rdoubledual}
\eea
with $E_{\abgd} $ being the Levi-Civita tensor. Here the convention adopted for the Levi-Civita symbol $\ep_{\abgd}$ is $\ep_{0123} = 1$.

This special property of the Gauss-Bonnet term is only true in the four-dimensional spacetime where the expression $g_{\mu\nu} E_{\abgd}$, when antisymmetrized over any five indices, vanishes. Then, the identity
\bea
g_\mn {\mathcal G} = -4 G_{\mu\be\ga\de} R_{\nu}^{\pt\mu \be\ga\de} ,
\eea 
can be proved, and the variation of ${\mathcal G}$ with respect to the metric can be calculated to obtain  
\bea
\de \left( \sqrt{-g} \, {\mathcal G} \right) = -4 \sqrt{-g} \, D_\ga \left( G_\al^{\pt\al\be\ga\de} \de \Ga^\al_{\pt\al \be\de} \right) ,
\eea
where $g$ is the determinant of the metric $g_\mn$, and $\Ga^\al_{\pt\al\be\de}$ are the Christoffel symbols with $D_\al$ being the covariant derivative associated with them. The above variation contributes as a surface term in the action, and hence is dropped in the equations of motion \cite{Lanczos:1938sf}. 

The futility of adding ${\mathcal G}$ multiplied by a constant to the action did not stop the interest in the Gauss-Bonnet term. On the contrary, it stimulated further investigations in dimensions higher than four \cite{Cai:2003gr, Konoplya:2004xx, Cai:2005ie, Camanho:2009vw}, motivated by string theory \cite{Zwiebach:1985uq, Boulware:1985wk}. Remarkably, a nontrivial theory in the four-dimensional spacetime was recently established by setting the constant in front of ${\mathcal G}$ proportional to $1/(D-4)$ with $D$ being the dimension of the spacetime \cite{Glavan:2019inb}. When taking the limit $D\rightarrow 4$, the Gauss-Bonnet term gives a finite contribution to the Einstein field equations and thus modifies GR.  

Another direction to investigate possible modifications to GR generated by the Gauss-Bonnet term is introducing a scalar field that couples to it \cite{Torii:2005xu, Nojiri:2005vv, Boulware:1986dr, Koivisto:2006ai, Cognola:2006sp, Leith:2007bu, Akbarieh:2021vhv}. The idea also originates from string theory where a scalar field, called the dilaton, appears when the extra dimensions are compactified to recover the metric tensor of the four-dimensional spacetime \cite{1974NuPhB..81..118S}. One well-studied case is when the Gauss-Bonnet term couples to an exponential function of the scalar field in the action, and inherits the name dilatonic Einstein-Gauss-Bonnet theory \cite{Kanti:1995vq, Torii:1996yi, Kanti:1997br, Torii:1998gm, Pani:2009wy, Kanti:2011jz, Ayzenberg:2014aka, Cunha:2016wzk, Ayzenberg:2018jip, Gonzalez:2018aky, Blazquez-Salcedo:2016enn, Konoplya:2019hml, Pierini:2021jxd, Yagi:2015oca, Witek:2018dmd, Wang:2021yll}; another basic case has the scalar field directly coupled to the Gauss-Bonnet term so that the field equations are invariant under a constant shift of the scalar field, and is given the name shift-symmetric Gauss-Bonnet theory \cite{Sotiriou:2013qea, Saffer:2021gak}. In these theories, the variation of the term involving ${\mathcal G}$ no long contributes as a surface term with the help of a nontrivial scalar field, and therefore modifies the Einstein field equations. Notice that the scalar field equations in these theories do not admit a constant scalar as the solution. So the metric solutions in these theories are generally different from that in GR. In this work, we are interested in another type of theories where a phenomenon called ``spontaneous scalarization'' can happen \cite{Damour:1993hw, Damour:1996ke, Harada:1998ge, Salgado:1998sg, Ramazanoglu:2016kul, Mendes:2016fby, Herdeiro:2018wub, Cunha:2019dwb, Antoniou:2017acq, Doneva:2017bvd, Silva:2017uqg, Doneva:2017duq, Kuan:2021lol}. This requires the theory to admit the trivial GR solution where the scalar field is a constant by assigning an even function of the scalar field coupled to ${\mathcal G}$ in the action. The theories in Refs.~\cite{Antoniou:2017acq, Doneva:2017bvd, Silva:2017uqg, Doneva:2017duq, Kuan:2021lol, Antoniou:2021zoy, Ventagli:2021ubn}, as proved therein, are such theories having solutions of spontaneous scalarization for both black holes and neutron stars (NSs). Here we refer them as scalar-Gauss-Bonnet theories of gravity with spontaneous scalarization, or simply scalar-Gauss-Bonnet theories.    

Theories in which compact objects can be scalarized are compelling, because they make identical predictions as those of GR in the weak-field regime, guaranteeing their validity as alternative theories of gravity, but are ready to be distinguished from GR via astrophysical observations of strong-field systems~\cite{Damour:1996ke, Wex:2014nva, Shao:2016ezh, Shao:2017gwu}. These observations include, for example, the shadow image of the supermassive black hole in the galaxy M87 obtained by the Event Horizon Telescope (EHT) \cite{EventHorizonTelescope:2019dse}, the orbital decay of binary pulsars observed by pulsar timing arrays \cite{Hulse:1974eb},
and the gravitational waves (GWs) from coalescences of compact binary systems detected by the LIGO and Virgo detectors \cite{LIGOScientific:2018mvr, LIGOScientific:2020ibl}.
The first comparison between the M87 shadow image and the numerically calculated black hole shadows in the typical scalar-Gauss-Bonnet theory proposed in Ref.~\cite{Doneva:2017duq} has been done by~\citet{Cunha:2019dwb}, where a preliminary constraint on the parameter of the theory is obtained. A recent work by \citet{Danchev:2021tew} first constrains the same theory using the orbital data of three binary pulsars under the framework of Bayesian analysis and MCMC strategy \cite{Shao:2017gwu}. As for using GW data to test scalar-Gauss-Bonnet theories, though it seems still early to talk about numerical bounds, the field is actively developing toward it. Particularly, \citet{Shiralilou:2020gah, Shiralilou:2021mfl} recently derives the post-Newtonian solution in scalar-Gauss-Bonnet theories to the first order beyond the quadrupole approximation, providing waveforms for the inspiralling stage of compact binary systems. Numerical simulations that required to build the waveforms for the merger stage of compact binary systems are unavailable but the investigation by \citet{Kovacs:2020pns} on the well-posedness of the initial-value problem in scalar-Gauss-Bonnet theories has set the first step in the direction.  
The ringdown stage of a coalescence can be approximated using perturbation theory in a single-body background spacetime, and the quasinormal modes (QNMs) of nonrotating black holes, which might serve as the basis for constructing the GWs emitted by the remnant black hole, have been studied in scalar-Gauss-Bonnet theories by~\citet{Bryant:2021xdh} and \citet{Staykov:2021dcj}. The QNMs of NSs, though in scalar-tensor theories without the Gauss-Bonnet term, have been studied by \citet{Mendes:2018qwo}.               

Our work aims at giving a detailed account of the spontaneously scalarized NSs in a typical scalar-Gauss-Bonnet theory with a massive scalar field, which complements the existing literature on spontaneously scalarized black holes as well as the work by~\citet{Doneva:2017duq} where NS solutions of spontaneous scalarization are obtained in a massless theory of such kind. In this paper, the study includes an analysis of the linearized scalar equation to find the onsets of spontaneous scalarization, illustrations of the numerical solutions to the group of nonlinear spherical equations, and a unified treatment of the slow-rotation and the tidal perturbations using the formalism of spherical decomposition introduced by~\citet{Regge:1957td} and developed for NSs in Refs.~\cite{1967ApJ...149..591T, 1969ApJ...155..163P, 1969ApJ...158....1T}. 
Our numerical solutions of the scalarized NSs are ready to use for constructing x-ray pulsar pulse profiles via the similar process demonstrated in Refs.~\cite{Silva:2018yxz, Xu:2020vbs, Hu:2021tyw} when adequate observational data, for example, from the NS Interior Composition Explorer (NICER) mission, are collected and analyzed to be compared with the theoretical predictions. The x-ray pulsar observations are also expected to measure the moment of inertia for NSs \cite{Hu:2020ubl, Miao:2021gmf}, providing another avenue to constrain the theory when compared to the numerical results for the moment of inertia calculated here. The tidal deformability, extracted from GW observations of NS mergers like GW170817 \cite{LIGOScientific:2017vwq}, provides a third way to test the theory, and certain values for the parameters of the theory have been excluded by a comparison between the measurement from GW170817 and the numerical results calculated here.

The paper is organized as follows. We start with a general setup for spherical NSs in scalar-Gauss-Bonnet theories in Sec.~\ref{sec:IIa}, and then numerically solve the linearized scalar equation in Sec.~\ref{sec:IIb} and the group of nonlinear equations in Sec.~\ref{sec:IIc} for the typical scalar-Gauss-Bonnet theory with a quadratic coupling function. Section~\ref{sec:III} deals with time-independent perturbations of the scalarized NSs. It starts with a brief review of the perturbation theory using spherical decomposition in Sec.~\ref{sec:IIIa}. Then numerical results for the tidal deformability and the moment of inertia of the scalarized NSs are presented in Sec.~\ref{sec:IIIb} and Sec.~\ref{sec:IIIc}. Finally, conclusions drawn from our study are summarized in Sec.~\ref{sec:sum}.  
The appendices provide detailed equations and analyses in our study. Appendix~\ref{app1} is about the expansion of the scalarized solution at the center of the star. Appendix~\ref{app2} is for the expansion of the scalarized solution at infinity. Appendix~\ref{app3} displays the radial equations for the even-parity perturbations with the angular momentum number $l \ge 2$. Appendix~\ref{app4} discusses a numerical issue in calculating the tidal deformability of the scalarized NSs. Appendix~\ref{app5} displays the radial equations for the odd-parity perturbations.

Throughout the work, equations are written in the geometrized unit system where $G=c=1$ unless conventional units appear in numerical results for the intuitive perception of the physical quantities. The sign convention of the metric is $(-, +, +, +)$.

\section{spherical NSs in scalar-Gauss-Bonnet gravity}
\label{sec:II}

\subsection{General setup}
\label{sec:IIa}
The action for the scalar-Gauss-Bonnet theory of gravitation to be studied has the general form \cite{Antoniou:2017acq, Doneva:2017bvd, Silva:2017uqg}
\bea
S &=& \frac{1}{16\pi} \int d^4 x \sqrt{-g} \left( R - g^\mn \prt_\mu \Ph \prt_\nu \Ph - U(\Ph) + F(\Ph) {\mathcal G} \right)
\nonumber \\
&& + S_m ,
\eea 
where the scalar field nonminimally couples to the metric through the product of the coupling function $F$ and the Gauss-Bonnet term ${\mathcal G}$. We assume that the scalar field $\Ph$ is absent in the matter action $S_m$. The field equations
\bea
G_\mn &=& 8\pi \left( T_m \right)_\mn + \left( T_\Ph \right)_\mn + \left( T_{\rm GB} \right)_\mn ,
\nonumber \\
D_\al D^\al \Ph &=& -\frac{1}{2} \frac{dF}{d\Ph} {\mathcal G} + \frac{1}{2} \frac{dU}{d\Ph},
\label{fieldequations}
\eea 
can be obtained by taking variations with respect to the metric and the scalar field. In the above equations, $G_\mn$ is the Einstein tensor, $\left( T_m \right)_\mn$ is the energy-momentum tensor for conventional matter, $\left( T_\Ph \right)_\mn$ is the energy-momentum tensor of a minimally coupled scalar
\bea
\left( T_\Ph \right)_\mn := \prt_\mu \Ph \prt_\nu \Ph - \frac{1}{2} g_\mn \left(g^{\al\be} \prt_\al \Ph \prt_\be \Ph + U \right) ,
\eea
and $\left( T_{\rm GB} \right)_\mn$ representing the contribution from the Gauss-Bonnet term  is 
\bea
\left( T_{\rm GB} \right)_\mn = 4 G_{\mu\al\nu\be}D^\al D^\be F ,
\eea
where $G_\abgd$ defined in Eq.~\rf{Rdoubledual} is the double dual of the Riemann tensor while $D_\al$ is the covariant derivative.  

Before we proceed to write down the ordinary differential equations (ODEs) using the spherical ansatz, let us mention that though the energy-momentum conservation equation for conventional matter 
\bea
D_\al \left(T_m \right)^{\al\be} = 0,
\label{coneq}
\eea
is merely a consequence of the field equations,\footnote{One can prove $D_\al \left( T_\Ph \right)^{\al\be} + D_\al \left( T_{\rm GB} \right)^{\al\be} = 0$ with the help of the scalar field equation.} components of it turn out to be handy to manipulate compared to the field equations themselves.

Now let us take the static spherical ansatz
\bea
ds^2 = -e^{2\nu} dt^2 + e^{2\mu} dr^2 + r^2 \left( d\th^2 + \sin^2\th d\ph^2 \right) ,
\eea 
and the perfect-fluid assumption for the conventional matter
\bea
\left( T_m \right)_{\al\be} = (\ep + p) u_\al u_\be + p g_{\al\be} ,
\eea
where the metric functions $\mu, \, \nu$, the fluid proper energy density $\ep$, and the fluid proper pressure $p$ are all functions of the radial coordinate $r$ only, while the 4-velocity field of the star is simply $u^\al = (e^{-\nu}, 0, 0, 0)$. Then, the $tt$- and $rr$-components of the modified Einstein field equations, the scalar field equation, and the $r$-component of the matter conservation equation conveniently give the group of ODEs for $\mu, \, \nu, \, \Ph$ and $p$:
\bea
\mu' &=& \frac{ - 2(e^{2 \mu}-1) + 16\pi r^2 e^{2 \mu} \ep +r^2 e^{2 \mu} U + r^2 \Ph^{\prime\, 2} }{4 \left( r + 2 \left(1-3e^{-2 \mu}\right)F' \right)} 
\nonumber \\
&& + \frac{ 8 \left(1-e^{-2 \mu}\right) F'' }{4 \left( r + 2 \left(1-3e^{-2 \mu}\right)F' \right)},
\nonumber \\
\nu' &=& \frac{ 2 (e^{2 \mu} - 1) + 16\pi r^2 e^{2 \mu} p - r^2 e^{2 \mu} U + r^2 \Ph^{\prime\, 2}}{4 \left(r + 2(1 - 3e^{-2 \mu} ) F' \right)} ,
\nonumber \\
\Ph'' &=& \left( \mu' - \nu' - \frac{2}{r} \right) \Ph' - \frac{1}{2} e^{2 \mu} {\mathcal G} \frac{dF}{d\Ph}  + \frac{1}{2} e^{2 \mu} \frac{dU}{d\Ph} ,
\nonumber \\
p' &=& -(\ep + p) \, \nu' ,
\label{odes}
\eea 
where the prime denotes the derivative with respect to $r$, and hence
\bea
F' = \frac{dF}{d\Ph} \Ph', 
~~
U' = \frac{dU}{d\Ph} \Ph', 
~~
F''= \frac{dF}{d\Ph} \Ph'' + \frac{d^2F}{d\Ph^2} \Ph^{\prime\, 2}  .
\eea
With the expression for the Gauss-Bonnet term
\bea
{\mathcal G} = \frac{8 e^{-2 \mu} \left[\left(1-3e^{-2 \mu}\right) \mu' \nu'-\left(1-e^{-2 \mu}\right) \left(\nu''+\nu^{\prime\,2}\right)\right]}{r^2},
\label{backgroundGB}
\eea
and an equation of state (EOS) that expresses $\ep$ in terms of $p$, the above ODEs can be cast into the standard form of $dy/dx = f(x,y)$ for numerical integration.

To conduct the numerical integration, we take the scalar potential $U$ as
\bea
U(\Ph) = \left(  \frac{2\pi}{\la_\Ph} \right)^2 \Ph^2 = \frac{1}{\lambdabar_\Ph^2 }  \Ph^2, 
\eea
to represent a massive scalar field whose mass is $m_\Ph = h/\la_\Ph$ with $\la_\Ph$ being its Compton wavelength and $h$ being the Planck constant. 
The coupling function $F$ in our work is assigned to be
\bea
F(\Ph) = \xi \Ph^2 ,
\eea 
with $\xi$ being a constant. It matches the coupling function in Ref.~\cite{Doneva:2017duq}
\bea
\la^2 f(\Ph) = \pm \frac{\la^2}{2\be} \left( e^{-\be \Ph^2} - 1 \right) ,
\eea
via $\xi = \mp \la^2 / 2$ when $\Ph$ is small.

In the rest of this section, we first use the method demonstrated by~\citet{Xu:2020vbs} to find the values of, say, the central energy density of the star, for the onsets of spontaneous scalarization. Then, full nonlinear solutions of spontaneous scalarization will be found around these critical values of the central energy density.  

\begin{figure}[h!]
  \includegraphics[width=\linewidth]{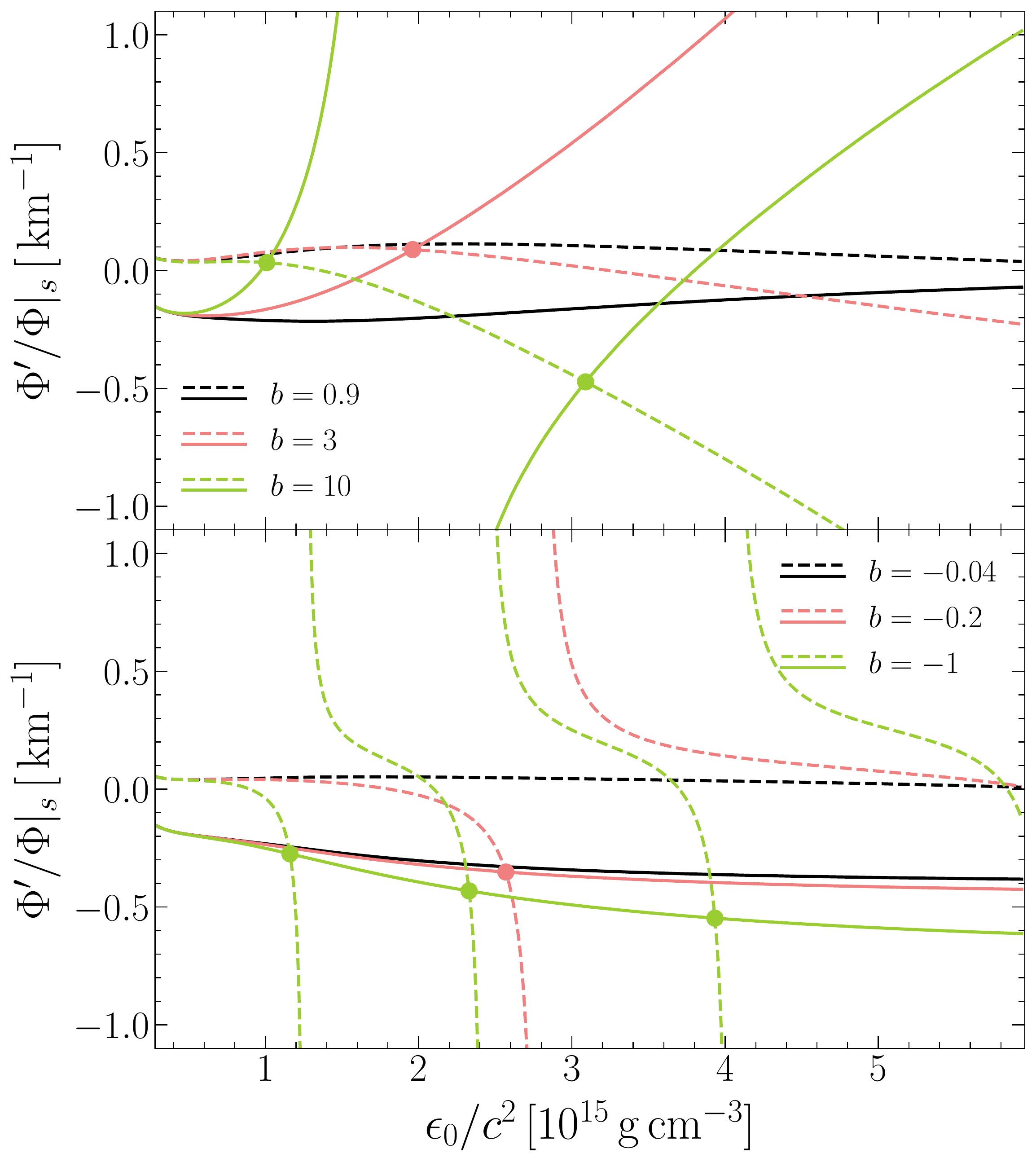}
  \caption{$\Phi' / \Phi \big|_s$ with respect to the central energy density
  $\ep_0$.  The dashed curves are from the interior solutions while the solid curves are from the exterior solutions. The value of $a$ is fixed to $1$ and the values of $b$ are shown in the panels; $a$ and $b$ are defined in Eqs.~\rf{eq:ab}. The EOS used is AP4. Notice that there are no intersections for $b = 0.9$ and $b=-0.04$.}  
 \label{fig_lin1}
\end{figure} 

\begin{figure}[h!]
  \includegraphics[width=\linewidth]{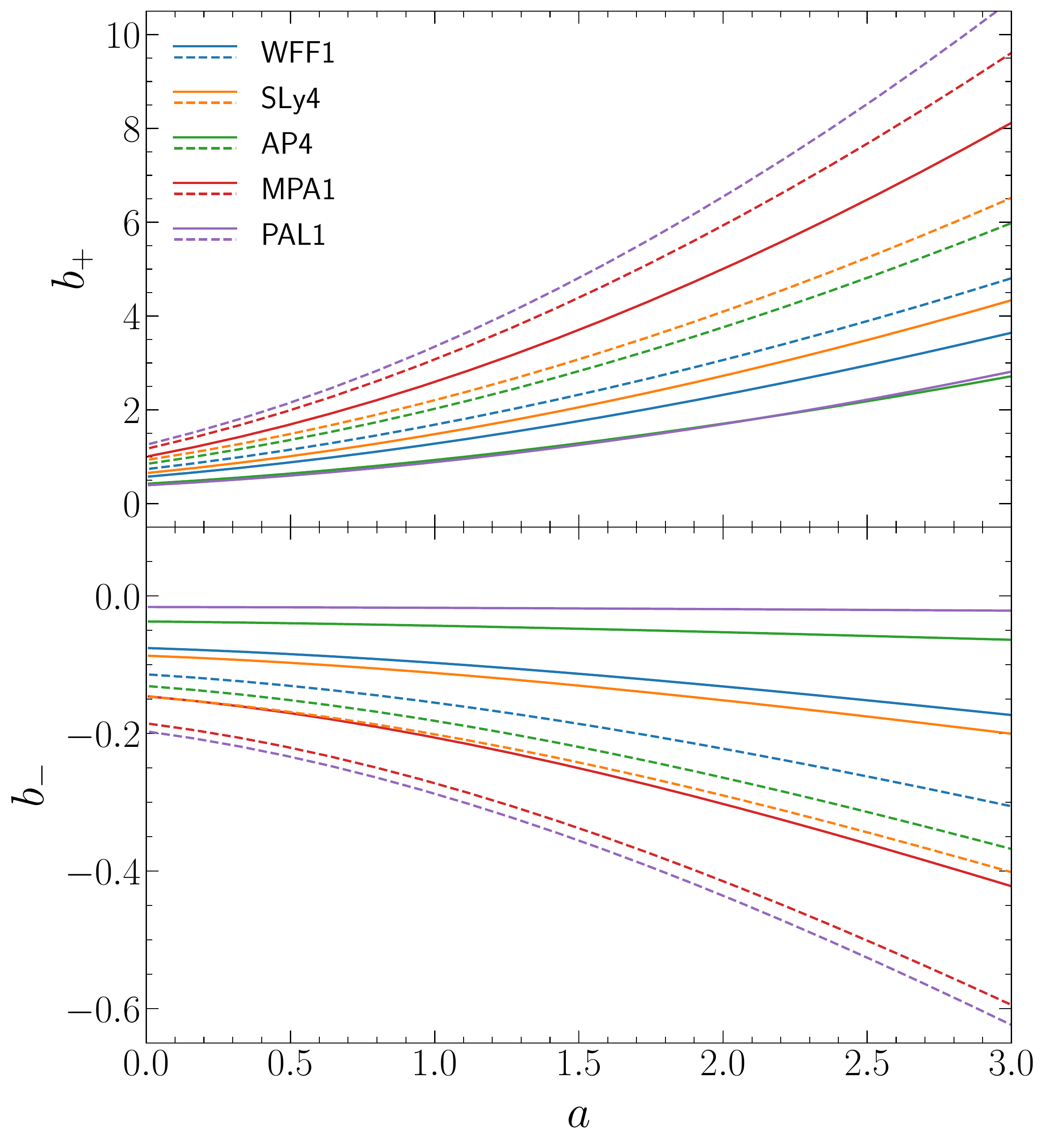}
  \caption{The boundary of the parameter space for spontaneous scalarization. The solid curves give the values of $a$ and $b$ for spontaneous scalarization to happen at the highest central energy density for a given EOS, while the dashed curves demand spontaneous scalarization to happen at the central energy density corresponding to the maximum mass of the NS for a given EOS.}
 \label{fig_lin2}
\end{figure}

\subsection{Parameter space and onsets of spontaneous scalarization }
\label{sec:IIb}

\begin{figure*}[h!]
 \includegraphics[width=0.9\linewidth]{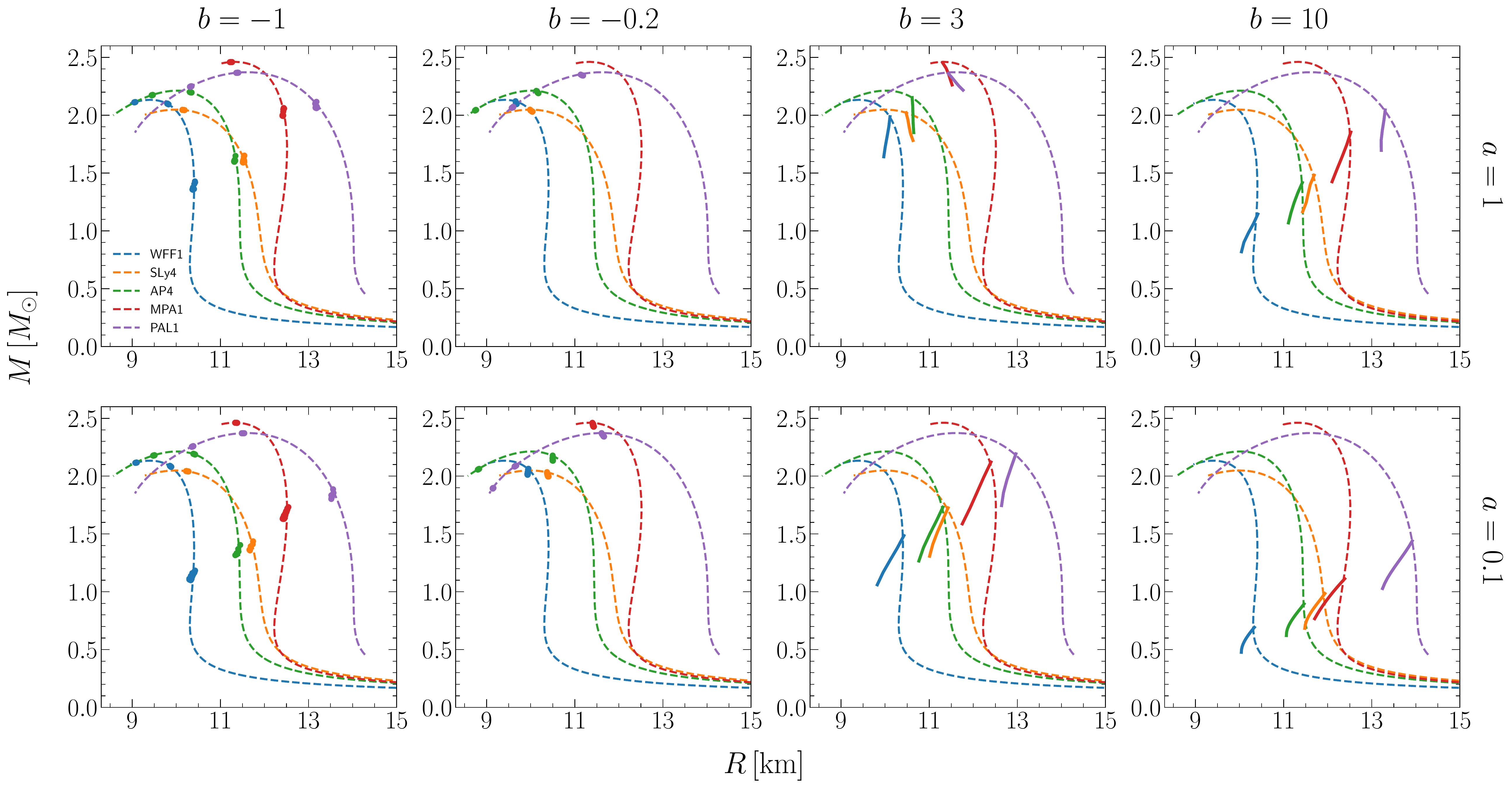}
 \caption{Mass-radius relation for NSs. The dashed curves are GR results, while the solid lines and the dots are the $M$-$R$ segments for the scalarized NSs. The same five EOSs as in Fig.~\ref{fig_lin2} are used: WFF1, SLy4, AP4, MPA1, and PAL1. Note that there are no scalarization solutions when $a=1,\, b=-0.2$ for the EOS MPA1 as its critical value $b_-$ at $a=1$ shown in Fig.~\ref{fig_lin2} is slightly less than $-0.2$.   }
\label{fig_nonlin_MR}
\end{figure*}

\begin{figure*}[h!]
 \includegraphics[width=0.9\linewidth]{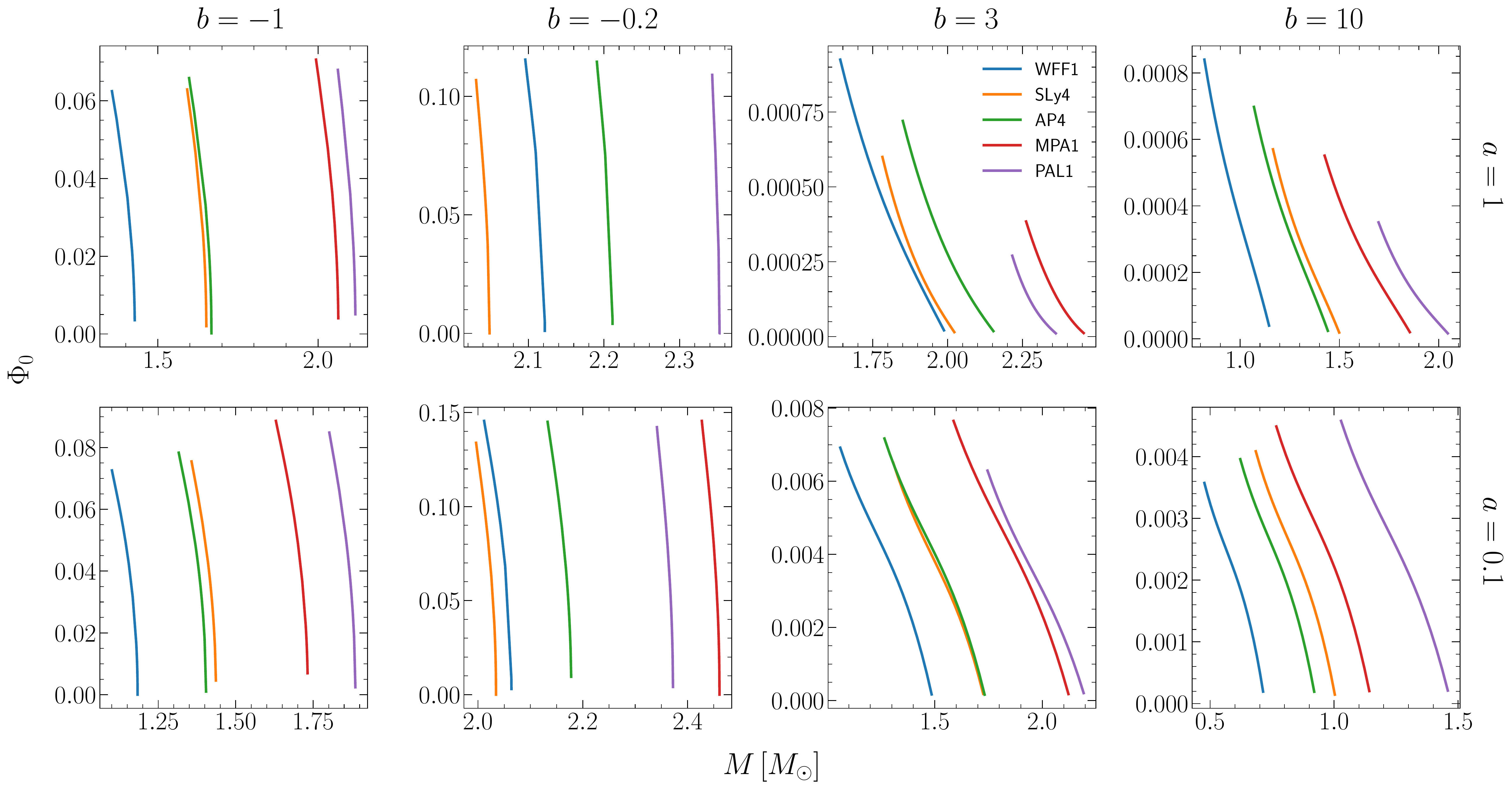}
 \caption{The scalar field at the center of the star versus the mass of the star for the solutions of spontaneous scalarization in Fig.~\ref{fig_nonlin_MR}. To keep the plots clear, only the solutions in the lowest mass band are plotted when there are more than one interval of spontaneous scalarization along the $M$-$R$ curve. The same set of colors as in Fig.~\ref{fig_nonlin_MR} is used for the five EOSs.     }
\label{fig_nonlin_Mbphc}
\end{figure*}

The set of the ODEs in Eqs.~\rf{odes} always admits the GR solutions for $\mu, \, \nu$ and $p$ together with a trivial scalar field $\Ph = 0$. The solutions of spontaneous scalarization where $\Ph$ is nontrivial however have requirements for both the star parameter which we take as the central energy density $\ep_0$, and the theory parameters $m_\Ph$ and $\xi$. We reveal the parameter space of $m_\Ph$ and $\xi$ for spontaneous scalarization as well as the critical values of the central energy density of the star for given values of $m_\Ph$ and $\xi$ by analyzing the linearized scalar field equation.

The linearized scalar field equation is the scalar field equation in Eqs.~\rf{odes} with the metric functions $\mu$ and $\nu$ taking their GR solution. It is solved in two regions: the interior and the exterior of the star. The interior scalar field is obtained by directly integrating from the center of the star where the value of $\Ph$ can be assigned arbitrarily as it only scales the solution and the value of $\Ph'$ is zero to keep $\Ph''$ finite. The exterior solution is computed by integrating from the surface of the star where the value of $\Ph$ can again be assigned arbitrarily but the value of $\Ph'$ needs to be deliberately selected to make sure that the asymptotically flat condition $\Ph \rightarrow 0$ at infinity is satisfied. Specifically speaking, a general exterior solution to the linearized scalar field equation has the asymptotical expression 
\bea
\Ph \rightarrow \Ph_+ e^{r/\lambdabar_\Ph} + \Ph_- e^{-r/\lambdabar_\Ph} ,
\label{phasy}
\eea   
where the functions $\Ph_+$ and $\Ph_-$ are finite when $r$ goes to infinity. The physical solution that $\Ph$ goes to zero at infinity excludes the $\Ph_+$ branch.

For calculating the interior solution, practical numerical integrations cannot begin exactly at the center of the star due to the $\propto 1/r$ factor in the scalar field equation. They begin at a radius very close to the center, and hence corrections to the values of $\Ph$ and $\Ph'$ at the starting point are necessary. The corrections can be approximated by the Taylor expansion of $\Ph$ at the center, whose expansion coefficient at the order of $r^2$ is displayed in Eq.~\rf{lincph} in Appendix~\ref{app1}. For calculating the exterior solution, the shooting strategy combined with the bisection method is used to find suitable values of $\Ph'$ at the surface of the star to make $\Ph$ vanish at infinity. The explicit expressions for the asymptotic functions $\Ph_+$ and $\Ph_-$ are irrelevant to our numerical approach, but for the completeness of the analysis as well as for illustrating the asymptotic behavior of the nonlinear solution, they are derived in Appendix~\ref{app2}.

Once the interior and the exterior solutions to the linearized scalar field equation are computed separately, the condition for them to form an entire solution is that the values of $\Ph'/\Ph$ at the surface of the star (denoted as $\Ph'/\Ph|_s$) given by the solutions on both sides match each other. For given metric functions $\mu$ and $\nu$ of a NS in GR, the match is not guaranteed. In other words, only specific NSs in GR possess solutions to the linearized scalar field equation. The parameters of these NSs, for example, their central energy densities or their masses, are critical in the sense that they mark possible onsets of spontaneous scalarization. To numerically find the critical values of the central energy density, we compute $\Ph'/\Ph|_s$ from the interior and the exterior solutions for different NSs in GR, and plot it versus the central energy densities of the NSs. When the curve of $\Ph'/\Ph|_s$ from the interior solution intersects the curve of $\Ph'/\Ph|_s$ from the exterior solution, we have a match and obtain a critical central energy density.

For the convenience in numerical calculation, we use $l_0 = 10\, {\rm km}$ as the length unit and parametrize other quantities using it in our work. Especially, the parametrized versions of $m_\Ph$ and $\xi$ are denoted as $a$ and $b$. In conventional units, $m_\Ph$ and $\xi$ in terms of $a$ and $b$ are
\bea
m_\Ph &=& \frac{ah}{2\pi l_0} = a \times 1.97 \times 10^{-11}\,{\rm eV} ,
\nonumber \\
\xi &=& b \, l_0^2  = b \times 10^8 \, {\rm m^2} .
\label{eq:ab}
\eea  
In Fig.~\ref{fig_lin1}, $\Ph'/\Ph|_s$ from the interior and the exterior solutions are plotted for changing values of the central energy density of the NS with $a=1$ and $b$ taking positive values $b=0.9,\, 3,\, 10$ and negative values $b=-0.04, \, -0.2, \,-1$. The metric of the NS in GR is solved using the EOS AP4. Four other EOSs, including WFF1, SLy4, MPA1, and PAL1, are verified to give similar results.  
The main observation from the plots is that the curve of $\Ph'/\Ph|_s$ from the interior solution and the curve of $\Ph'/\Ph|_s$ from the exterior solution start to intersect at the highest central energy density when $|b|$ is sufficiently large. As $|b|$ increases, the existing intersection moves to left and a new intersection comes out again at the highest central energy density. The procedure repeats as long as $|b|$ continuously increases. So the larger $|b|$ is, the more intersections accumulate.

For given values of $a$, the values of $|b|$ when the first intersection just appears define the boundary of the theory parameter space for spontaneous scalarization. The solid curves in Fig.~\ref{fig_lin2} depict the numerical results of the boundary when different EOSs are used, namely that the value of $b$ needs to be either greater than $b_+$ or less than $b_-$ for spontaneous scalarization to happen when $a$ is given. As the first intersection appears at the highest central energy density and then moves to left when $|b|$ increases, we can ask what the value of $|b|$ is when the intersection goes down to the central energy density that corresponds to the maximum mass of the NS for the EOS used. This value of $|b|$ gives a naive estimate for the boundary of the theory parameter space if the spontaneous scalarization is to be stable. The dashed curves in Fig.~\ref{fig_lin2} show the results for the five different EOSs. For a given value of $a$, if the value of $b$ is between the solid curve and the dashed curve for an EOS in consideration, then there are spontaneously scalarized NSs in the theory though they are expected to be unstable due to their extremely high central energy densities.


\subsection{Solutions of spontaneously scalarized NSs }
\label{sec:IIc}

\begin{figure}[h!]
 \includegraphics[width=0.8\linewidth]{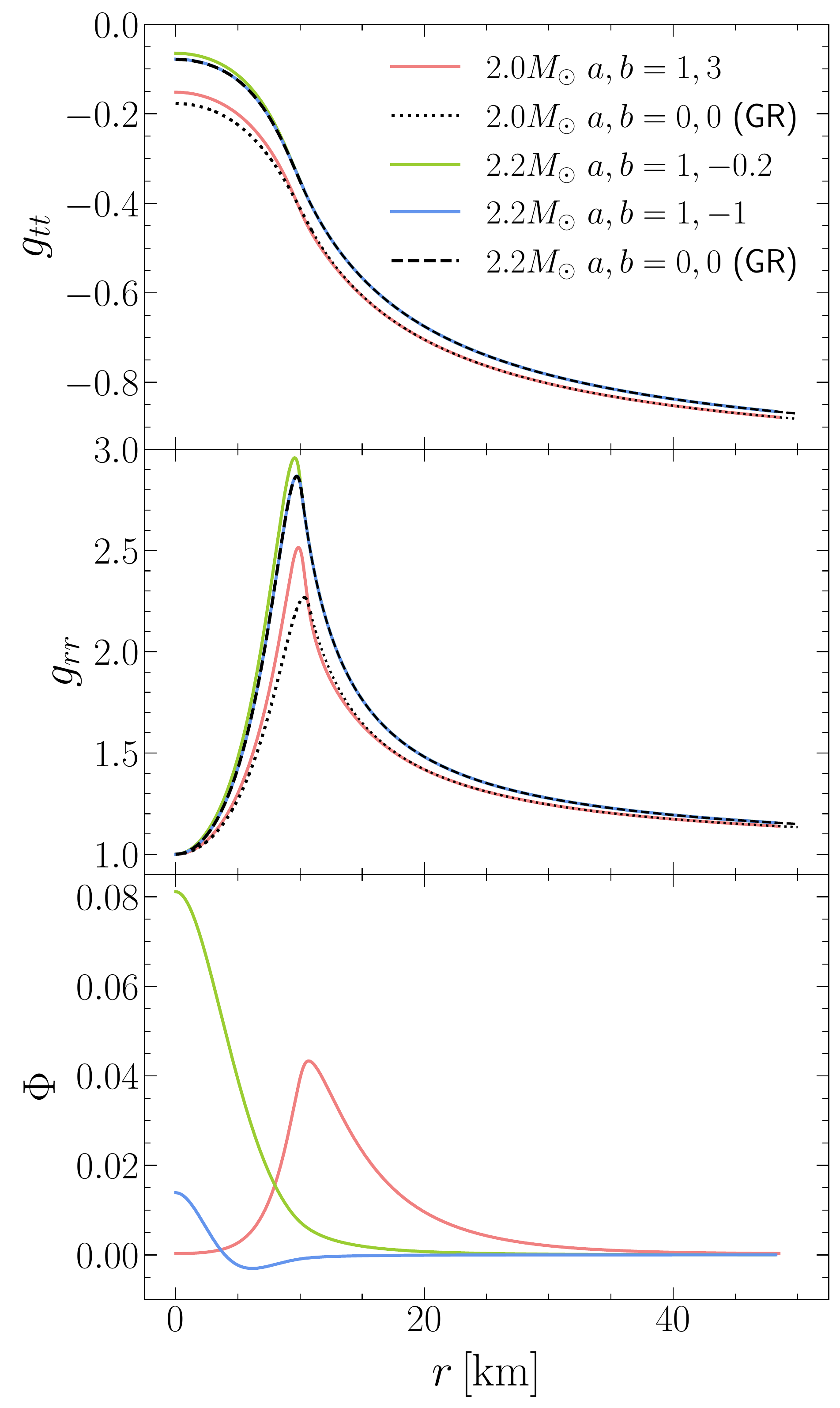}
 \caption{The metric components and the scalar field of three representative scalarized NSs. The dashed and dotted curves show the metric functions of GR NSs with the same masses for comparison. Notice the difference between the profiles of the scalar field for $b>0$ and $b<0$. Also note that the scalar field of the scalarized NS with $2.2\,M_{\odot}$ in the case of $a, b=1, -1$ has a node, as it belongs to the second scalarization band in the $M$-$R$ diagram.}
\label{fig_nonlin_ind}
\end{figure}

For values of $m_\Ph$ and $\xi$ in the parameter space for spontaneous scalarization, solutions of spontaneously scalarized NSs can be searched around the critical central energy densities given by the linearized analysis described above. Noticing that only $\mu,\, \Ph$ and $p$ are coupled in the group of the ODEs in Eqs.~\rf{odes}, the search can be conducted through adjusting the value of $\Ph$ at the center of the star (denoted as $\Ph_0$) while the value of $\mu$ at the center of the star is set to zero and the value of $p$ at the center is given by the EOS once the central energy density $\ep_0$ is set to a value close to one of the critical values obtained in the linearized analysis. Similarly to the linearized case, the numerical integrations have to start at a tiny but nonzero radius. So corrections to the starting values of $\mu,\, \Ph$ and $p$ that can be approximated by their Taylor expansions at the center of the star are requisite. Appendix~\ref{app1} discusses the expansions and gives the equations from which the expansion coefficients at the order of $r^2$ can be numerically solved.

Differently from the linearized case, now it is unnecessary to integrate the interior and the exterior regions separately. The integration goes directly from the center of the star to a large enough radius, where $\Ph$ likely blows up due to the same $e^{r/\lambdabar_\Ph}$ factor unless the central value $\Ph_0$ is properly chosen so that the $e^{r/\lambdabar_\Ph}$ branch of the solution vanishes. By applying the shooting strategy to find out the appropriate values of $\Ph_0$ for valid values of $\ep_0$, solutions of spontaneous scalarization with $\Ph \rightarrow 0$ at infinity are attained.

Representative numerical results are plotted on the mass-radius relation diagram in Fig.~\ref{fig_nonlin_MR}. The first thing to be noticed is that solutions of spontaneous scalarization are found around every critical central energy density given by the linearized analysis for the case of $\xi < 0$ while for the case of $\xi > 0$, only the critical central energy density corresponding to the first intersection in the $\Ph'/\Ph|_s$ plot (see Fig.~\ref{fig_lin1}) is an onset of spontaneous scalarization. Detailed numerical integrations show that around the critical central energy densities corresponding to the second and higher intersections in the $\Ph'/\Ph|_s$ plot for the case of $\xi>0$, solutions of spontaneous scalarization fail to develop because of an unexpected divergence soon after the integrations start. The divergence can only originate from the high nonlinearity in Eqs.~\rf{odes}.  

\def\arraystretch{1.3}
\begin{table*}
\caption{Numerical properties of the five representative scalarized neutron stars plotted in Fig.~\ref{fig_nonlin_ind}.   }
\begin{tabular}{l c c c c c | c c | c c }
  \hline   \hline
 \multicolumn{6}{p{10cm}|}{ \hspace{1.65cm}  Parameters} & \multicolumn{2}{p{2.5cm}|}{ \hspace{0.2cm} ISCO properties}  & \multicolumn{2}{p{3.8cm}}{ \hspace{0.2cm} perturbation properties}   \\
   $a, b$ & $\Ph_0$ & $\ep_0/c^2$ ($10^{15}\, {\rm g/cm^3}$) & $M\, (M_{\odot})$ & $R$ (km) &  Baryonic mass$\,(M_{\odot})$ &  $r$ (km) & $\frac{d\ph}{dt}\, ({\rm ms}^{-1})$ & $\lambda/M^5$ & $I$ ($10^{45}\, {\rm g\,cm^2}$) \\
  \hline
   $0, 0$\, (GR) & 0 & 0.9833 & 1.400 & 11.42 & 1.571 & 12.40 & 9.867 & $2.62 \times 10^2$ & 1.324 \\
   $0.1, -1$ & $1.752\times 10^{-2}$ & 0.9860 & 1.400 & 11.45 & 1.571 & 12.35 & 9.929 & $2.77 \times 10^2$ & 1.323 \\ 
   $0.1,3$   & $5.237\times 10^{-3}$ & 1.083 & 1.400 & 10.89 & 1.569 & 12.41 & 9.571 & $1.86 \times 10^3$ & 1.241 \\
  \hline
   $0, 0$\, (GR) & 0 & 1.532 & 2.000 & 11.00 & 2.392 & 17.72 & 6.907 & $1.64 \times 10^1$ & 2.152\\
   $1, 3$ & $2.741 \times 10^{-4}$ & 1.682 & 2.000 & 10.63 & 2.390 & 16.90 & 7.959 & $7.35 \times 10^1$ & 2.067\\
  \hline   
   $0, 0$\, (GR) & 0 & 2.308 & 2.200 & 10.31 & 2.706 & 19.49 & 6.279 & 3.62 & 2.309\\
   $1, -0.2$ & $8.116\times 10^{-2}$ & 2.570 & 2.200 & 10.15 & 2.706 & 19.50 & 6.822 & 4.42 & 2.260\\
   $1, -1$ & $1.388\times 10^{-2}$ & 2.319 & 2.200 & 10.32 & 2.706 & 19.69 & 6.717 & 5.75 & 2.315\\ 
  \hline
\end{tabular}
\label{tab1}
\end{table*}

Secondly, one cannot help noticing the very short ranges on the $M$-$R$ diagram for the solutions of spontaneous scalarization. This is especially true for the case of $\xi<0$ where for each interval of spontaneous scalarization the mass of the NS only changes less than $0.1\,M_\odot$, making the solutions actually difficult to find without the hint of the critical central energy densities from the linearized analysis. Our numerical results also show that increasing the value of $|\xi|$ does not widen the mass intervals of spontaneous scalarization. The fact that the occurrence of spontaneous scalarization is limited to NSs in the narrow mass bands is a consequence of the quadratic coupling function chosen. Using a different type of coupling function, for example the exponential one studied by~\citet{Doneva:2017duq}, the intervals of spontaneous scalarization can be much wider.

Thirdly, we notice that all the solutions of spontaneous scalarization distribute themselves on the left side of the $M$-$R$ curves of GR for the case of $\xi>0$, implying that these scalarized NSs are smaller than the NSs in GR with the same mass. The situation for the case of $\xi < 0$ is a little subtle as there can be several scalarization bands along the $M$-$R$ curve of GR and the scalarization bands are extremely narrow. A zooming-in observation on them shows that the $M$-$R$ segments of the scalarized NSs can actually depart the $M$-$R$ curves of GR on their right sides and then turn left and cross the $M$-$R$ curves of GR when $\xi$ is sufficiently negative. In this case, a scalarized NS can have a radius smaller than, equal to, or greater than that of a NS in GR with the same mass, depending on where it sits along the $M$-$R$ segment. Though it is theoretically interesting to capture the details, it is unlikely useful in testing the theories with negative $\xi$ in observations because the deviations from GR are anyhow too small.

Fourthly, a noticeable feature, at least for the case of $\xi>0$, is that the $M$-$R$ segments of the scalarized NSs have only one end attached to the $M$-$R$ curves of GR. This suggests that the $M$-$R$ curves bifurcate at the onsets of spontaneous scalarization, and that the branches of spontaneous scalarization develop as the scalar field at the center of the star increases to finite values where the solutions of spontaneous scalarization cease to exist. Plots of the scalar field at the center of the star versus the mass of the star, as shown in Fig.~\ref{fig_nonlin_Mbphc}, help to illustrate the existence and the evolution of spontaneous scalarization. Starting at the onsets of spontaneous scalarization where the central scalar $\Ph_0$ is technically zero, the solutions evolve to reach smaller NS masses as $\Ph_0$ increases to finite values and stops. Naturally the value of $\Ph_0$ should drop back to zero after it reaches a maximum. The reason for the abrupt cutoff is that the Taylor expansion coefficients $\mu_2, \, \nu_2$ and $\Ph_2$ solved from Eqs.~\rf{2ndrec} start to have imaginary parts when $\Ph_0$ and $\ep_0$ happen to be in certain regions. Noticing that it is because of the quartic nature of the recurrence equations that they admit complex solutions for $\mu_2, \, \nu_2$ and $\Ph_2$, we can draw the conclusion that the high nonlinearity in Eqs.~\rf{odes} again restricts the existence of spontaneous scalarization. A similar restriction for spontaneously scalarized black holes in scalar-Gauss-Bonnet theories, expressed in an analytical inequality, can be found in Refs.~\cite{Antoniou:2017acq, Doneva:2017bvd, Silva:2017uqg}.  
 
Lastly, a careful comparison of the 8 panels in Fig.~\ref{fig_nonlin_MR} tells us that as $m_\Ph$ increases or $|\xi|$ decreases, the intervals of spontaneous scalarization shift to the higher end along the $M$-$R$ curves of GR, and vice versa.

From Fig.~\ref{fig_nonlin_MR} we see that the two main parameters of the scalarized NSs, namely their masses and radii, barely deviate from the values of NSs in GR for the case of $\xi<0$. This makes distinguishing those scalarized NSs from NSs in GR using observations very difficult. In fact, even for the case of $\xi > 0$, the largest deviation in radius between a scalarized NS and a NS in GR with the same mass is only about $1\, {\rm km}$ as shown in Fig.~\ref{fig_nonlin_MR}. This just lies within the uncertainties of current measurements of NS radius \cite{LIGOScientific:2018cki, Riley:2019yda, Miller:2019cac, Riley:2021pdl, Miller:2021qha}. The degeneracy is even obvious when individual solutions are compared. In Fig.~\ref{fig_nonlin_ind} we plot the metric components of three representative scalarized NSs calculated using the EOS AP4 to compare with the metric components of the NSs in GR with the same EOS and masses. To conclude our study of the spherical solution of spontaneously scalarized NSs, certain parameters of five representative scalarized NSs are displayed in Table~\ref{tab1} for reference.

\section{Time-independent perturbations}
\label{sec:III}

In this section, we deal with a slowly rotating and tidally deformed NS under spontaneous scalarization using perturbation theory. The formalism that decomposes the metric perturbation into spherical harmonics introduced by \citet{Regge:1957td} and later developed for NSs by \citet{1967ApJ...149..591T} is adopted. We first give a brief review of the formalism.

\subsection{Review of the perturbation theory in spherical background }
\label{sec:IIIa}
Spherical harmonics form complete sets of orthogonal functions on a 2-sphere, and therefore are naturally the bases to expand the angular dependence of perturbations around a spherical background. Depending on the rotational properties of the quantities on the 2-sphere, scalar, vector, and tensor spherical harmonics may be used. Following the definitions of \citet{Thorne:1980ru}, the transverse vector spherical harmonics and the traceless transverse tensor spherical harmonics with rank-2 used here are constructed from the usual scalar spherical harmonics $Y_{lm}$ via, 
\bw
\bea
\boldsymbol{Y}^E_{lm} & := & \frac{1}{ \sqrt{l(l+1)} } r \boldsymbol{\nabla} Y_{lm} 
= \frac{r}{ \sqrt{l(l+1)} } \left( {\boldsymbol{d\th}} \,\prt_\th + {\boldsymbol{d\ph}} \, \prt_\ph \right) Y_{lm} ,
\nonumber \\
\boldsymbol{Y}^B_{lm} & := & \frac{1}{ \sqrt{l(l+1)} } \boldsymbol{r} \times \boldsymbol{\nabla} Y_{lm} 
= \frac{r}{ \sqrt{l(l+1)} } \left( - \frac{ {\boldsymbol{d\th}} }{\sin\th}  \,\prt_\ph + {\boldsymbol{d\ph}} \, \sin\th \, \prt_\th \right) Y_{lm} ,
\nonumber \\
\boldsymbol{T}^{E2}_{lm} & := & \sqrt{ \frac{2}{(l-1)(l+2)} } \left[ r \boldsymbol{\nabla} \boldsymbol{Y}^E_{lm} \right]^{\rm ST} 
\nonumber \\
&=& N_l \, r^2 \left[{\boldsymbol{d\th}}\otimes{\boldsymbol{d\th}} \, \prt_\th^2 + {\boldsymbol{d\ph}}\otimes{\boldsymbol{d\ph}} \left( \sin\th\cos\th \, \prt_\th + \prt_\ph^2 \right) + \left( {\boldsymbol{d\th}}\otimes{\boldsymbol{d\ph}} 
 + {\boldsymbol{d\ph}}\otimes{\boldsymbol{d\th}} \right) \prt_\ph \left(\prt_\th - \cot\th \right)  \right]^{\rm T} Y_{lm} ,
\nonumber \\
\boldsymbol{T}^{B2}_{lm} & := & \sqrt{ \frac{2}{(l-1)(l+2)} } \left[ r \boldsymbol{\nabla} \boldsymbol{Y}^B_{lm} \right]^{\rm ST}    
\nonumber \\
&=& N_l \, r^2 \left[ \left( - \frac{{\boldsymbol{d\th}}\otimes{\boldsymbol{d\th}}}{\sin\th} + {\boldsymbol{d\ph}}\otimes{\boldsymbol{d\ph}} \, \sin\th \right) \prt_\ph \left( \prt_\th - \cot\th \right) + {\boldsymbol{d\th}}\otimes{\boldsymbol{d\ph}} \, \sin\th \, \prt_\th^2 - {\boldsymbol{d\ph}}\otimes{\boldsymbol{d\th}} \left( \cos\th \, \prt_\th + \frac{\prt_\ph^2}{\sin\th} \right)  \right]^{\rm S} Y_{lm} ,
\label{vtsphhar}
\eea    
\ew
where $\boldsymbol{\nabla}$ is the gradient in three-dimensional Euclidean space, $N_l := \sqrt{2/\left[(l-1)l(l+1)(l+2)\right]}$, and $\{ {\boldsymbol{dr}}, \, {\boldsymbol{d\th}}, \, {\boldsymbol{d\ph}} \}$ is used as the basis for covariant components of a vector in the spherical coordinates $(r, \th, \ph)$. The superscript ``S'' means symmetrizing the tensor and the superscript ``T'' means removing the trace. By dropping the factor $N_l\, r^2$, the covariant components of $\boldsymbol{T}^{E2}_{lm}$ without removing the trace and the covariant components of $\boldsymbol{T}^{B2}_{lm}$ can be recognized as $\psi_{l{\pt m}\mn}^{{\pt l}m}$ and $\chi_{l{\pt m}\mn}^{{\pt l}m}$ used by \citet{Regge:1957td}. 

The perturbative quantities to be expanded using the spherical harmonics are the metric perturbation $\de g_{\mn}$, the fluid displacement $\xi_\mu$ and the scalar perturbation $\de \Ph$. On the 2-sphere, the metric perturbation can be decomposed as four scalars:
\bea
\de g_{tt}, \ \de g_{tr}, \ \de g_{rr}, \ \de g_{\th\th} +   \frac{1}{ \sin^2\th} \de g_{\ph\ph},
\nonumber 
\eea
two vectors:
\bea
 \de g_{t\th} \, {\boldsymbol{d\th}} + \de g_{r\ph} \, {\boldsymbol{d\ph}}, \ \de g_{r\th} \, {\boldsymbol{d\th}} + \de g_{r\ph}\, {\boldsymbol{d\ph}},
 \nonumber
\eea
and a traceless tensor:
\bea
\Big[\de g_{\th\th} \, {\boldsymbol{d\th}}\otimes{\boldsymbol{d\th}} + \de g_{\th\ph} \left({\boldsymbol{d\th}}\otimes{\boldsymbol{d\ph}} + {\boldsymbol{d\ph}}\otimes{\boldsymbol{d\th}} \right) + \de g_{\ph\ph}\,{\boldsymbol{d\ph}}\otimes{\boldsymbol{d\ph}} \Big]^{\rm T} ,
\nonumber
\eea   
while the fluid displacement can be decomposed as two scalars:
\bea
\xi_t, \ \xi_r,
\nonumber
\eea
and a vector: 
\bea
\xi_\th \, {\boldsymbol{d\th}} + \xi_\ph {\boldsymbol{d\ph}} .
\nonumber
\eea
Furthermore, the expansions of all the 11 quantities can be conveniently divided into two parts according to the parity of the spherical harmonics: the so-called even-parity ones, $\{ Y_{lm}, \boldsymbol{Y}^{E}_{lm}, \boldsymbol{T}^{E2}_{lm} \}$, that pick up a sign of $(-1)^l$ under a parity transformation, and the so-called odd-parity ones, $\{ \boldsymbol{Y}^{B}_{lm}, \boldsymbol{T}^{B2}_{lm} \}$, that pick up a sign of $(-1)^{l+1}$ under the same transformation. Because the perturbative field equations directly inherit the parity properties of $\de g_\mn$ and $\xi_\mu$, perturbations with different types of parity do not mix. The same conclusion is also true for the angular momentum numbers $l$ and $m$, allowing us to consider perturbations with specific values of $l$ and $m$ rather than a sum over all the values of them.

With the separation of even and odd parities, the 11 quantities become 22. But since there is no scalar spherical harmonics of odd parity, 7 of them trivially vanish. Moreover, the freedom to choose coordinates at the order of the perturbation can help us eliminate another 4 of them. Using the Regge-Wheeler gauge \cite{Regge:1957td}, the four quantities eliminated are all in the metric perturbation: the two vectors with even parity and the traceless tensor with even parity and with odd parity. This finally leaves us $11$ nonvanishing quantities to express in terms of the spherical harmonics of a specific parity as well as specific $l$ and $m$, namely, for even parity
\bea
\de g_{tt} = e^{2\nu} H_0 Y_{lm} , 
\quad
\de g_{tr} &=& H_1 Y_{lm} ,
\quad
\de g_{rr} = e^{2\mu} H_2 Y_{lm}
\nonumber \\
\de g_{\th\th} + \frac{1}{\sin^2\th} \de g_{\ph\ph}  &=& 2 r^2 K Y_{lm} ,
\nonumber \\
\xi_t = W_0 Y_{lm} , 
\quad
\xi_r &=& W_1 Y_{lm}
\nonumber \\
\xi_\th \, {\boldsymbol{d\th}} + \xi_\ph \, {\boldsymbol{d\ph}} &=& V \boldsymbol{Y}^{E}_{lm} \frac{\sqrt{l(l+1)}}{r} ,
\nonumber \\
\de \Ph &=& X Y_{lm} ,
\label{evenper}
\eea
and for odd parity
\bea
\de g_{t\th} \, {\boldsymbol{d\th}} + \de g_{t\ph} \, {\boldsymbol{d\ph}} &=& h_0 \boldsymbol{Y}^{B}_{lm} \frac{\sqrt{l(l+1)}}{r} ,
\nonumber \\
\de g_{r\th} \, {\boldsymbol{d\th}} + \de g_{r\ph} \, {\boldsymbol{d\th}} &=& h_1 \boldsymbol{Y}^{B}_{lm} \frac{\sqrt{l(l+1)}}{r} ,
\nonumber \\
\xi_\th \, {\boldsymbol{d\th}} + \xi_\ph \, {\boldsymbol{d\ph}} &=& h_2 \boldsymbol{Y}^{B}_{lm} \frac{\sqrt{l(l+1)}}{r} ,
\label{oddper}
\eea
where $H_0, \, H_1, \, H_2, \, K, \, W_0, \, W_1,\, V,\, X$ and $h_0, \, h_1, \, h_2$ are unknown functions of $t$ and $r$ to be solved from the perturbative field equations. 

To obtain differential equations for the 11 perturbative functions using the field equations \rf{fieldequations}, the fluid displacement $\xi_\mu$ needs to be related to the perturbation of the 4-velocity, the perturbation of the energy density, and the perturbation of the pressure. The relations are given, for example, by~\citet{Gittins:2020mll}, as
\bea
\De u^\mu &=& \frac{1}{2} u^\mu u^\al u^\be \De g_{\al\be}, 
\nonumber \\
\frac{\De n}{n} &=& -\frac{1}{2} \left( g^{\al\be} + u^\al u^\be \right) \De g_{\al\be},
\label{perun}
\eea   
for the Lagrangian perturbation of the 4-velocity $\De u^\mu$ and the Lagrangian perturbation of the particle number density $\De n$ in terms of the Lagrangian perturbation of the metric $\De g_{\al\be}$. The fluid displacement comes into Eqs.~\rf{perun} when we change the Lagrangian perturbations to the Eulerian perturbations via the definition $\De := \de + {\mathcal L}_\xi$, where ${\mathcal L}_\xi$ is the Lie derivative along the fluid displacement. Therefore, the Eulerian perturbation of the 4-velocity to be used in the perturbative field equations is   
\bea
\de u^\mu &=& \frac{1}{2} u^\mu u^\al u^\be \left( \de g_{\al\be} + D_\al \xi_\be + D_\be \xi_\al \right) 
\nonumber \\
&& - \left( \xi^\la D_\la u^\mu - u^\la D_\la \xi^\mu \right) .
\label{peru}
\eea
The Eulerian perturbations of the energy density and the pressure can be calculated from their Lagrangian perturbations
\bea
\De \ep = \left( \ep + p \right) \frac{\De n}{n}, 
\quad
\De p = \ga p \frac{\De n}{n} ,
\label{perepp}
\eea
where the first equation is a result of the first law of thermodynamics under the assumption that the fluid elements are adiabatic, and the second equation can be regarded as the definition of the adiabatic index $\ga$. Given an EOS relating $\ep$ and $p$, the adiabatic index $\ga$ can be obtained via
\bea
\ga = \frac{\ep+p}{p} \frac{dp}{d\ep} .
\eea   

Now substituting $\de g_{\al\be}$ and $\xi_\mu$ from Eqs.~\rf{evenper} and \rf{oddper}, as well as the zeroth-order quantities, into Eqs.~\rf{perun}, \rf{peru} and \rf{perepp}, we can obtain the spherical decompositions for the perturbations of the 4-velocity, the energy density, and the pressure. The perturbative field equations then can be obtained. Once the angular dependence is detached, they give equations for $H_0, \, H_1, \, H_2, \, K, \, W_0, \, W_1,\, V,\, X$ and $h_0, \, h_1, \, h_2$. In Appendices~\ref{app3} and \ref{app5}, these equations are displayed with the time dependence set to $e^{i\si t}$, where the constant $\si$ stands for the frequency of a normal or quasinormal mode of the star.

\begin{figure*}
 \includegraphics[width=0.9\linewidth]{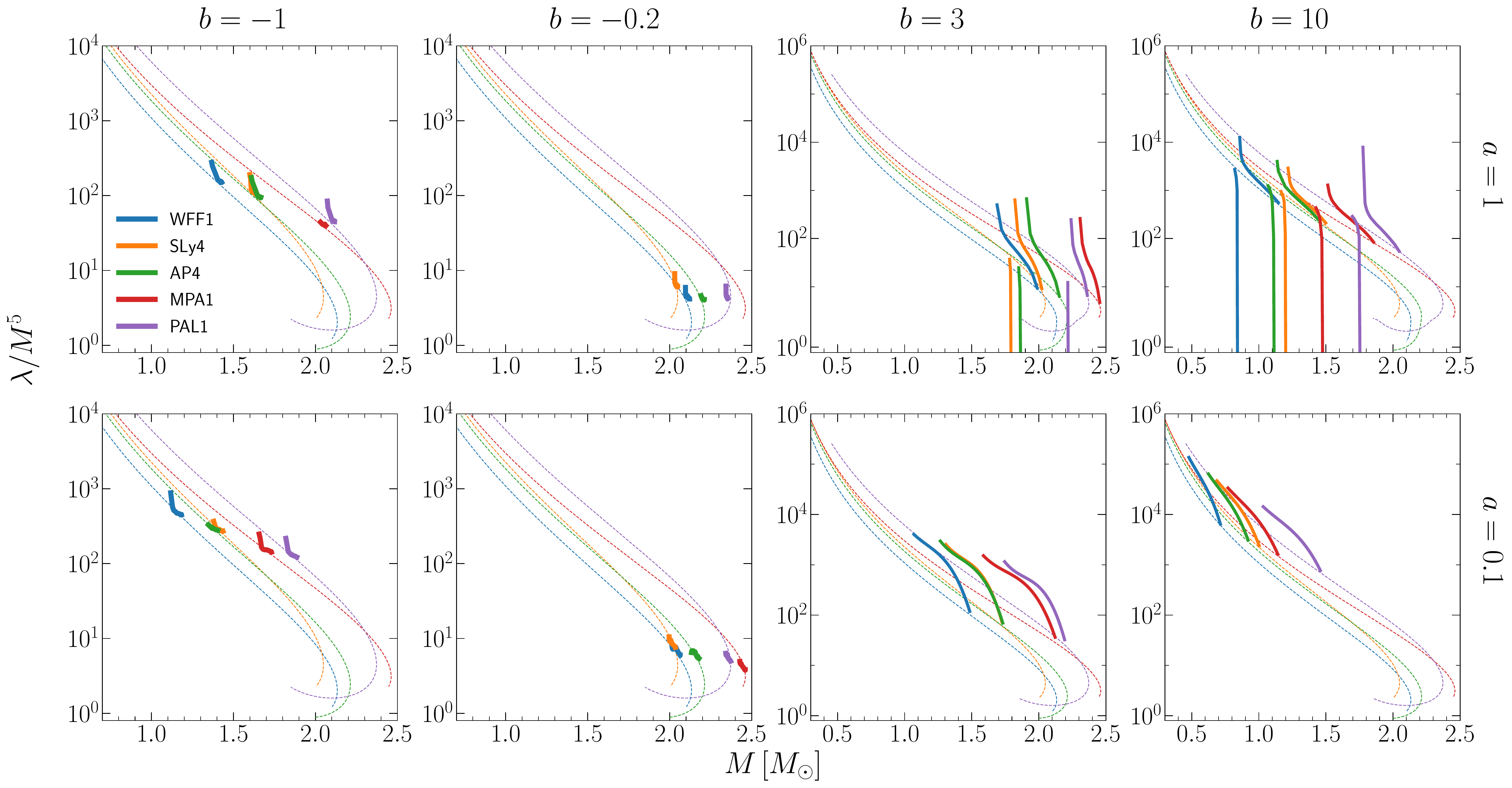}
 \caption{The dimensionless tidal deformability $\lambda/M^5$ versus the mass of NSs. The dashed curves are GR results, while the solid curves are results of the scalarized NSs in the lowest mass bands in Fig.~\ref{fig_nonlin_MR}. Please notice the discontinuities in the panels of $a = 1, b = 3, 10$. Details are discussed in the text.   }
\label{fig_nonlin_Mtidal}
\end{figure*}

\subsection{Tidal deformation: static even-parity $l=2$ }
\label{sec:IIIb}
Our first application of the perturbation theory in spherical background is calculating the perturbation of the metric under a static external tidal field. The equations and the numerical approach to solve them are summarized in Appendix~\ref{app3}. The quantity to be calculated is the tidal deformability $\la$, defined via \cite{Hinderer:2007mb}
\bea
Q_{2m} = -\la {\mathcal E}_{2m} ,
\eea 
where $Q_{2m}$ with $m=0, \pm 1, \pm 2$ are the induced quadrupole moments of an originally spherical star and ${\mathcal E}_{2m}$ are the spherical components of the static external tidal field. Both $Q_{2m}$ and ${\mathcal E}_{2m}$ are defined in the asymptotic expansion of the metric perturbation $\de g_{tt}$ of even parity and angular momentum number $l=2$, namely
\bea
\de g_{tt} &=& e^{2\nu} H_0 \sum\limits_{m=-2}^2 A_m Y_{2m} 
\nonumber \\
& \rightarrow&  \frac{3}{r^3} \sum\limits_{m=-2}^2 Q_{2m} Y_{2m} - r^2 \sum\limits_{m=-2}^2 {\mathcal E}_{2m} Y_{2m} ,
\label{perH0}
\eea 
where $A_m$ are $m$-related constants to give angular dependence that matches the second line. Because the perturbative function $H_0$ is a solution to Eqs.~\rf{evenodes} that are independent of $m$, its leading asymptotic terms
\bea
H_0 \rightarrow  \frac{Q_{2m}}{A_m} \frac{3}{r^3} - \frac{ {\mathcal E}_{2m}}{A_m} r^2  ,
\label{asyh0tidal}
\eea 
read off from Eq.~\rf{perH0}, must be $m$-independent.
Therefore, the tidal deformability $\la = - Q_{2m}/{\mathcal E}_{2m}$ does not depend on $m$.

To calculate $H_0$ and extract its leading asymptotic expansion coefficients as shown in Eq.~\rf{asyh0tidal}, we set $\si = 0$ in Eqs.~\rf{evenodes} so they eventually simplify to three coupled ODEs for $H_0, \, K$ and $X$ that can be numerically integrated from the center of the star. As usual, practical integrations start at a tiny radius rather than exactly at the center of the star to avoid numerical zeros in denominators. The values of $H_0, \, K, \, X$ and $X'$ at the tiny radius where the integrations start are then necessary inputs in the integrator. They are given by the Taylor expansions in Eqs.~\rf{evencbc} with two of the expansion coefficients, say $H_{00}$ and $X_0$ as used in Eqs.~\rf{evencbc0}, free to choose for the moment. The two coefficients are actually not free as soon as the physical asymptotic behavior is required for the solution. Similarly to the spherical solution, the asymptotic expression of a general solution to the perturbative equations contains both the factors of $e^{-r/\lambdabar_\Ph}$ and the factors of $e^{r/\lambdabar_\Ph}$, while the contributions from the latter is supposed to be eliminated in the physical solution by adjusting the inputs of the integration through $H_{00}$ and $X_0$. The problem of generating the physical solution is much simplified since the perturbative equations are homogeneous in the perturbative functions. Following \citet{Pani:2014jra}, the physical solution can be found by linearly combining any two independent solutions to make $X$ vanish at infinity instead of employing the shooting strategy that aims at finding suitable values of $H_{00}$ and $X_0$.

In our numerical approach, two independent solutions are found with $(H_{00}, \, X_0) = (1, \, 0 )$ and $( H_{00}, \, X_0 ) = ( 1, 1 )$, and then after the physical solution is constructed from them, two methods for computing the tidal deformability are available. The first one is extracting the coefficients $Q_{2m}/A_m$ and ${\mathcal E}_{2m}/A_m$ shown in Eq.~\rf{asyh0tidal} by directly fitting the numerical solution of $H_0$ in the asymptotic region and then computing $\lambda$. The second one is using the formula developed by~\citet{Hinderer:2007mb} with all the quantities defined on the surface of the star therein substituted by the corresponding values at a sufficiently large $r$ where the terms containing the factors of $e^{-r/\lambdabar_\Ph}$ can be safely dropped so the expansion of $H_0$ is exactly the same as that in GR for the formula to apply \cite{Hu:2021tyw}. The two methods are no doubt theoretically equivalent but the second one turns out to be too sensitive to numerical errors when $\lambdabar_\Ph$ is large. The first method is adopted in this work. The issue of the second method with large $\lambdabar_\Ph$ is discussed in detail in Appendix~\ref{app4}.    

Figure \ref{fig_nonlin_Mtidal} shows the results of the tidal deformability for the NSs in Fig.~\ref{fig_nonlin_MR}. For $\xi < 0$, the deviations are anyhow tiny so we only plotted the results for the scalarized NSs in the lowest mass bands. The results for $\xi > 0$ however are quite appealing. First and foremost, the scalarized NSs can have the dimensionless tidal deformability $\lambda/M^5$ an order of magnitude larger than that of the NSs in GR. As a matter of fact, it is counterintuitive that the tidal deformability of the scalarized NSs is larger than that of the NSs in GR at all since according to the $M$-$R$ diagram in Fig.~\ref{fig_nonlin_MR}, these scalarized NSs are more compact than their GR counterparts. Secondly, the plots for $a=1$ show jumps where the sign of the tidal deformability changes, suggesting that the tidal deformability of the scalarized NSs approaches infinities near the jumps. By investigating the numerically fitted expansion coefficients of $H_{00}$, we find that the coefficient in front of $r^2$ decreases to zero for the special scalarized NSs that have masses at the discontinuities in the plots for $a=1, \, b=3, 10$ in Fig.~\ref{fig_nonlin_Mtidal}. For scalarized NSs near the special masses, the sign of the $r^2$-coefficient in the asymptotic expansion of $H_{00}$ changes. This feature neither shows up in GR nor in the nonminimal scalar-tensor theories studied before \cite{Hu:2021tyw}. It is a qualitative difference between the scalar-Gauss-Bonnet gravity studied here and GR.     

The large deviations in the tidal deformability of the scalarized NSs from their GR counterparts are no doubt most helpful in testing the scalar-Gauss-Bonnet gravity in astrophysical observations. Simply taking the result from the observation of GW170817: $\lambda/M^5 \lesssim 720 $ for $M\approx 1.4M_{\odot}$ \cite{LIGOScientific:2018hze}, we can already exclude the point $(a, \, b)= (0.1, \,3)$ in the parameter space for the two popular EOSs AP4 and SLy4. To rule out a region in the parameter space of $(a,\, b)$, numerical results of the tidal deformability with $a$ and $b$ taking values near the point $(0.1, 3)$ are necessary to compare with the observational constraint. The implementation of the comparison algorithm into numerical code is beyond the scope of the present work but certainly worth further researching as more and more constraints on the tidal deformability of NSs are expected from future GW observations~\cite{KAGRA:2013rdx}.

\subsection{Slow rotation: static odd-parity $l=1$  }
\label{sec:IIIc}

\begin{figure*}
 \includegraphics[width=0.9\linewidth]{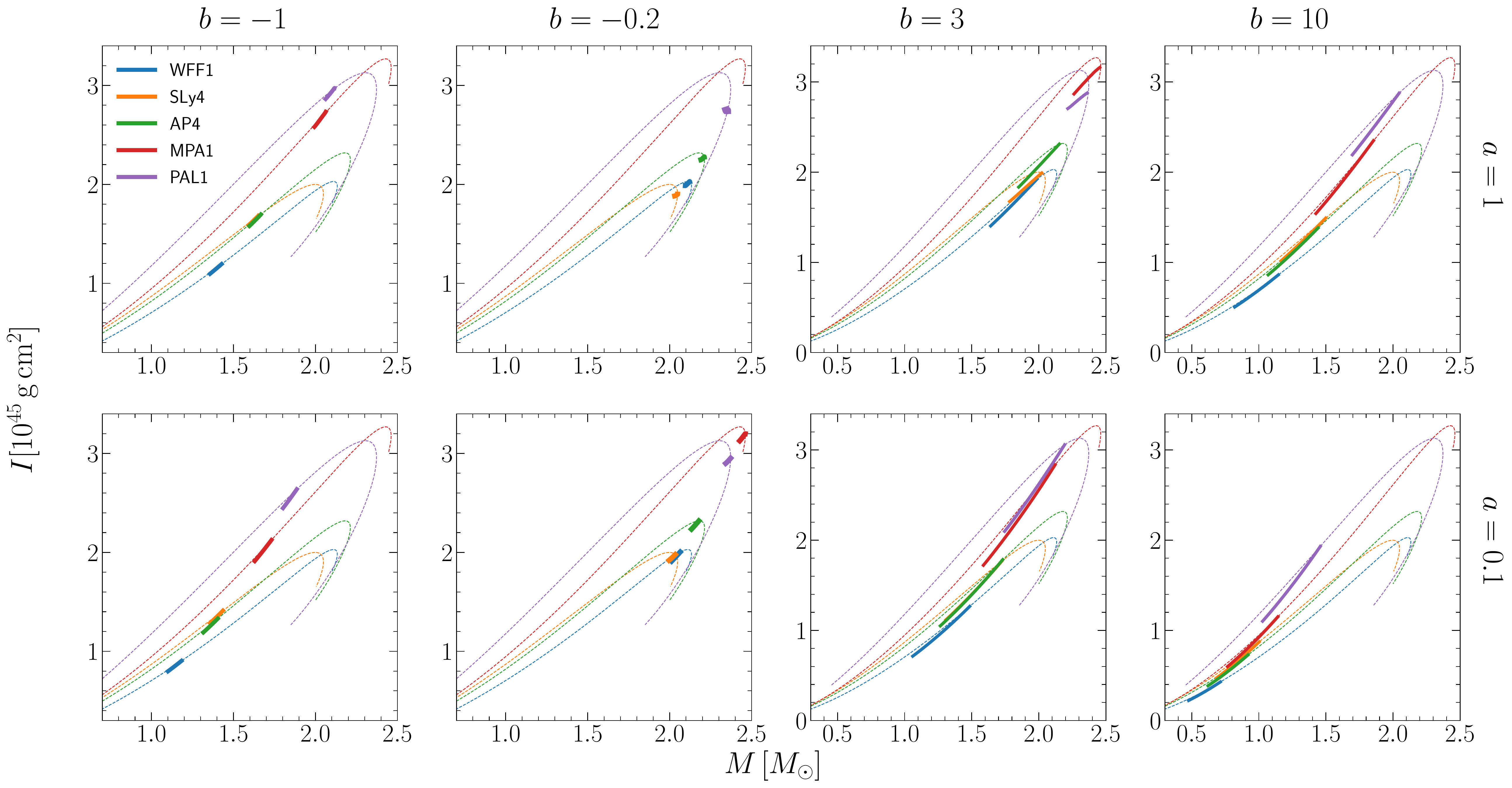}
 \caption{The moment of inertia versus the mass of NSs. The dashed curves are GR results, while the solid segments are results of the scalarized NSs in the lowest mass bands in Fig.~\ref{fig_nonlin_MR}.    }
\label{fig_nonlin_MI}
\end{figure*}  

The second application of the perturbation theory in spherical background is calculating the moment of inertia of slowly rotating scalarized NSs. The problem boils down to solving the first equation in Eqs.~\rf{oddodes} with $h_1 = 0$ and $\si \rightarrow 0$ while the product $\si h_2$ keeps finite and represents the radial part of the angular velocity of the star via the match \rf{slowrot}. Following~\citet{Damour:1996ke}, we assume the angular velocity $\Om$ to be constant so that the moment of inertia of the star can be defined as
\bea
I  := \frac{J}{\Om} ,
\eea 
where the angular momentum $J$ is extracted from the leading asymptotic term of $g_{t\ph}$. The match \rf{slowrot} indicates $l=1$ for a constant $\Om$, and $i\si h_2$ is to be substituted by $ -\sqrt{3/(4\pi) }\, \Om r^2$ in the first equation in Eqs.~\rf{oddodes}, which becomes a second-order ODE with a source term for $h_0$.

The numerical method to solve $h_0$ is integrating from the center of the star where $h_0$ has the expansion
\bea
h_0 = h_{00} r^2 + h_{02} r^4 + \cdots \,,
\eea 
when $l=1$. 
The leading coefficient $h_{00}$ can be set to 1 and the next-to-leading coefficient $h_{02}$ can be calculated using Eq.~\rf{oddstaticcbc} for given values of $\Om$. For an arbitrary value of $\Om$, the solution of $h_0$ includes an asymptotic term proportional to $r^2$ as well as an asymptotic term proportional to $1/r$. The former needs to be eliminated by elaborately choosing a proper value of $\Om$ so that the metric component $g_{t\ph}$ has the desired asymptotic behavior
\bea
g_{t\ph} = h_0 \sin\th \, \prt_\th Y_{10} &=& - \sqrt{ \frac{3}{4\pi} } \sin^2\th \, h_0
\nonumber \\ 
& \rightarrow&  -\frac{2J}{r} \sin^2\th ,
\label{gtphl1}
\eea
from which the angular momentum $J$ can be read off.

Before showing the numerical results, a comment on the equivalence between the approach used here and the approach used by \citet{Damour:1996ke} might be worthwhile. The metric ansatz $g_{t\ph} = \left(\om - \Om \right) r^2 \sin^2\th$ can be compared with Eq.~\rf{gtphl1} to find $h_0 = -\sqrt{4\pi/3} \left( \om - \Om\right) r^2$. Therefore solving $h_0$ is equivalent to solving $\om$ only that the constant angular velocity of the star $\Om$ appears as an input when solving $h_0$, while it becomes an output together with the angular momentum $J$ in the asymptotic expansion of $\om$,
\bea
\om \rightarrow \Om - \frac{2J}{r^3} + O\left( r^{-4} \right)  .
\eea
We have to admit that though we derived the equation for $h_0$, it is converted to the equation for $\om$ to solve numerically. Because there is no adjustable input in solving $\om$, it is practically more convenient than solving $h_0$.

\begin{figure*}
 \includegraphics[width=0.9\linewidth]{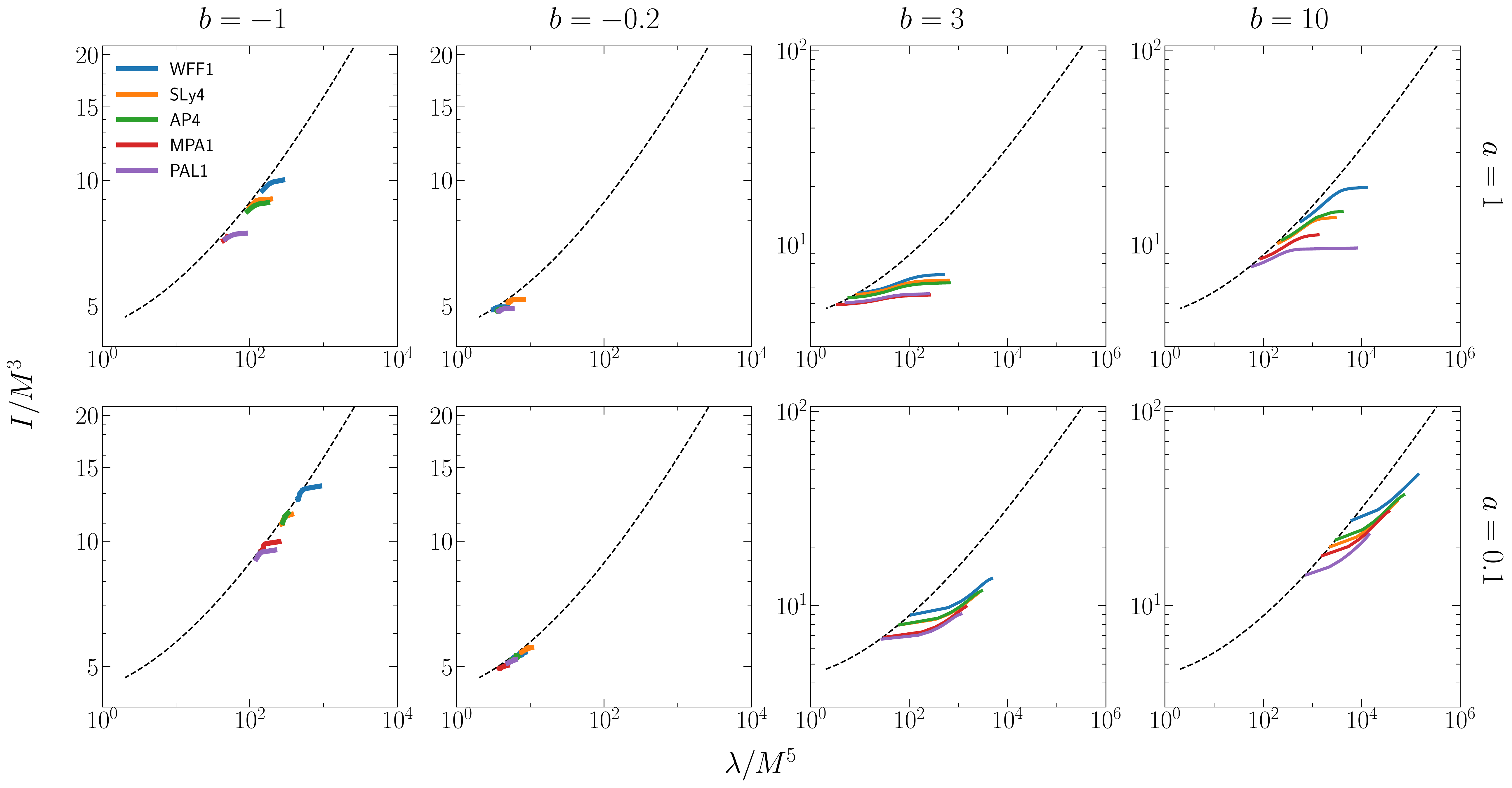}
 \caption{Plots for $I/M^3$ versus $\lambda/M^5$. The dashed curve is the numerically fitted universal relation in GR \cite{Yagi:2016bkt}, while the solid curves are results of the scalarized NSs in the lowest mass bands in Fig.~\ref{fig_nonlin_MR}. The panels of $a=1, \, b=3, 10$ only plot values of $\lambda$ on the right side of the discontinuities shown in Fig.~\ref{fig_nonlin_Mtidal}.   }
\label{fig_nonlin_Itidal}
\end{figure*}

In Fig.~\ref{fig_nonlin_MI}, the moment of inertia of the scalarized NSs in the lowest mass bands in Fig.~\ref{fig_nonlin_MR} is compared with that of NSs in GR. We see that the scalarized ones all have slightly smaller values of the moment of inertia compared with NSs in GR when the masses are the same, intuitively agreeing with the fact that the radii of most of them are smaller than those of their GR counterparts.\footnote{However, it also raises a puzzle for the remaining of them that have larger radii than their GR counterparts.}

With the tidal deformability and the moment of inertia calculated, we are ready to check if the universal relation between the dimensionless quantities $\bar \lambda := \lambda/M^5$ and $\bar I := I/M^3$ discovered by \citet{Yagi:2016bkt} in GR  applies to the scalar-Gauss-Bonnet theory studied here. Since Fig.~\ref{fig_nonlin_Mtidal} shows noticeable difference between the tidal deformability of the scalarized NSs and the tidal deformability of the NSs in GR for $\xi > 0$ while the difference in the moment of inertia shown in Fig.~\ref{fig_nonlin_MI} is very restrained, we expect that at least for $\xi>0$, the scalarized NSs have different $\bar I$-$\bar \lambda$ relations rather than the same universal relation as in GR. The plots in Fig.~\ref{fig_nonlin_Itidal} verify the expectation. The $\bar I$-$\bar \lambda$ segments corresponding to the scalarized NSs for $\xi > 0$ visibly deviate from the universal relation in GR. Moreover, the $\bar I$-$\bar \lambda$ segments for different EOSs do not overlap, so they fail to form a universal relation that is independent of EOSs. This could have implication for observations.
Universal relations involving quantities derived from higher-order calcualtion in the spin, e.g. the quadrupole moment and the shape parameter (cf. Ref.~\cite{Gao:2021uus}), are beyond the scope of this study.

\section{Summary}
\label{sec:sum}
In this work which complements the study of spontaneous scalarization of NSs in the massless scalar-Gauss-Bonnet theory of gravitation by \citet{Doneva:2017duq}, the same phenomenon is investigated in detail for the massive scalar-Gauss-Bonnet theory. A linearized analysis on the scalar field equation is conducted to discover the parameter space as well as the onsets of spontaneous scalarization first. Then, numerical solutions of spontaneous scalarization are computed and their properties are discussed. To gain more insights into the scalarized NSs, their static perturbative properties are investigated. These properties include the tidal deformability and the moment of inertia, both of which we treat under the unified framework of spherical decomposition.        

To summarize the results, we find that our solutions share similar qualitative characteristics with the solutions in the massless theory. First, solutions of spontaneous scalarization exist for both positive and negative coupling constants. Second, the scalar field at the center of the star emerges from zero at the onsets of spontaneous scalarization, and develops to finite values while the mass of the star changes (see Fig.~\ref{fig_nonlin_Mbphc}). But it does not drop back to zero again, namely that the onsets of spontaneous scalarization obtained by the linearized analysis are one-sided. Solutions of spontaneous scalarization cease to exist on the other end due to the nonlinearity of the leading-order recurrence equations \rf{2ndrec} at the center of the star. Equations \rf{2ndrec} only have real solutions for a certain range of the central parameters $\ep_0, p_0$ and $\Ph_0$, restricting the possible combinations of the star mass $M$ and the central scalar field $\Ph_0$ for solutions of spontaneous scalarization. Third, for negative coupling constants, spontaneous scalarization can happen in more than one interval along the $M$-$R$ curve (see Fig.~\ref{fig_nonlin_MR}). The scalarization interval corresponding to the lowest mass band has scalar solutions possessing zero node. The scalar solutions in the second lowest mass band possess one node, and so on.    

More importantly, the quantitative details of our results show differences from those in Ref.~\cite{Doneva:2017duq} since we use a different type of coupling function and consider a massive scalar field. The coupling function used here is quadratic in $\Ph$, which is intuitively weaker than the exponential coupling function used by~\citet{Doneva:2017duq}. Therefore, the ranges of the star mass for spontaneous scalarization in our theory are narrow compared to the results in Ref.~\cite{Doneva:2017duq}. We notice that increasing the absolute value of the coupling constant $\xi$ in our theory does not lead to wider bands of the star mass for spontaneous scalarization, indicating that the widths of the mass bands depend on the type of coupling function rather than the magnitude of the coupling constant. Because of the smallness of the spontaneous scalarization bands, we have to zoom in on their $M$-$R$ diagrams to observe so that we find that the $M$-$R$ segments of the scalarized NSs can cross the $M$-$R$ curves in GR when the coupling constant $\xi$ is sufficiently negative, a fact that might be easily ignored if the $M$-$R$ segments of the scalarized NSs spread out longer.              

The mass of the scalar field, according to Fig.~\ref{fig_nonlin_MR}, drives the spontaneously scalarized NSs to the higher end along the  $M$-$R$ curves in GR. Apart from that, its effect is mainly the Yukawa suppression factor $e^{-r/\lambdabar_\Ph}$ that once dropped, makes the asymptotic behavior of the scalarized solutions identical to that of the solutions in GR. In case the exponential terms are desired, for example, when the reduced Compton wavelength of the scalar field $\lambdabar_\Ph$ is much larger than the radius of the region that one is interested in, we employ the idea of double expansion, namely expanding the asymptotic solution in factors of $M/r$ and factors of $e^{-r/\lambdabar_\Ph}$, to come up with the asymptotic expressions \rf{nonlinasysol} for the first time to our knowledge. Unexpectedly, we find that Eqs.~\rf{nonlinasysol} do not have limits when $\lambdabar_\Ph \rightarrow \infty$, invalidating our attempt of matching them to the asymptotic expansions \rf{linasysol} in the massless scalar-tensor theory.   

The scalarized NSs found in our work show very moderate deviations from NSs in GR on the $M$-$R$ diagram, making the theory very easy to escape constraints from astrophysical observations of the mass and radius of NSs, especially with the uncertainty from various EOSs. Therefore, we further calculated their time-independent perturbative properties, including the tidal deformability and the moment of inertia. While the moment of inertia of the scalarized NSs only has insignificant differences from that of NSs in GR, remarkable deviations in the tidal deformability of the scalarized NSs from their GR counterparts are found for the case of a positive coupling constant. Not only can the magnitude of the tidal deformability of these scalarized NSs be significantly higher than that of the NSs in GR, but also the sign of the tidal deformability changes to negative where a discontinuity pops up at a special NS mass when the mass of the scalar field in the theory is large (see the panels of $a=1,\,b=3,10$ in Fig.~\ref{fig_nonlin_Mtidal}). The discontinuity is caused by the vanishing of the $r^2$-coefficient in the asymptotic expansion of the metric perturbation $\de g_{tt}$, a feature that does not occur in GR.

Distinguishing GR and the scalar-Gauss-Bonnet theory with a positive coupling constant becomes feasible with the discrepancies in the tidal deformability of the NSs. For the scalar-Gauss-Bonnet theory with a negative coupling constant, the quasinormal modes of the scalarized NSs might shed light on the way to distinguish them from NSs in GR using observations, which will be studied in a follow-up work. In addition, as the moment of inertia barely shows differences between the scalarized NSs and the NSs in GR, the renowned universal relation between the dimensionless tidal deformability $\lambda/M^5$ and the dimensionless moment of inertia $I/M^3$ discovered by \citet{Yagi:2016bkt} in GR no longer holds for the scalarized NSs (see Fig.~\ref{fig_nonlin_Itidal}).    
The tidal deformability of NSs has begun to be accessible in the gravitational wave data by the LIGO-Virgo-KAGRA Collaboration \cite{LIGOScientific:2017vwq, LIGOScientific:2018hze, KAGRA:2013rdx}, and the moment of inertia of binary pulsars is expected to be extractable via pulsar timing in the coming era of the Square Kilometre Array (SKA) \cite{Hu:2020ubl, Shao:2014wja, Weltman:2018zrl}. Once they are measured for a sufficient group of NSs with necessary precision, we might be able to tell whether GR should be regarded as an approximation of a fundamentally motivated scalar-Gauss-Bonnet theory \cite{Zwiebach:1985uq, Boulware:1985wk, 1974NuPhB..81..118S}, or at least, to exclude an appreciable region in the parameter space of it.

\acknowledgments 

We are grateful to the anonymous referee for the useful suggestions and comments, and Zexin Hu for helpful discussions.
This work was supported by the National SKA Program of China (2020SKA0120300), the National Natural
Science Foundation of China (11975027, 11991053, 11721303), 
the Young Elite Scientists Sponsorship Program by the China Association for
Science and Technology (2018QNRC001), the Max Planck Partner Group
Program funded by the Max Planck Society, and the High-performance Computing
Platform of Peking University. RX is supported by the Boya Postdoctoral Fellowship at Peking University.

\appendix
\section{Expansion of the spherical solution at the center of the star}
\label{app1}
To investigate the behavior of the solutions to the set of ODEs in Eqs.~\rf{odes} at the center of the star, the Taylor expansions 
\bea
\mu &=& \sum\limits_{n=0}^{\infty} \mu_n r^n ,
\quad
\nu = \sum\limits_{n=0}^{\infty} \nu_n r^n ,
\quad
\Ph = \sum\limits_{n=0}^{\infty} \Ph_n r^n ,
\nonumber \\
p &=& \sum\limits_{n=0}^{\infty} p_n r^n ,
\quad
\ep = \sum\limits_{n=0}^{\infty} \ep_n r^n ,
\label{cexp}
\eea
are substituted into Eqs.~\rf{odes} to find the recurrence relations for the expansion coefficients. 
We are only interested in the physical solutions that are smooth at $r=0$, namely those with 
\bea
\mu_1 = \nu_1 = \Ph_1 = p_1 = \ep_1 = 0 .
\eea
Then, the leading-order recurrence relations fix $\mu_0$ to zero while leaving $p_0$ and $\Ph_0$ free, from which nonvanishing $\mu_n, \, \nu_n, \, \Ph_n$ and $p_n$ for even $n$'s are generated by the higher-order recurrence relations. Note that the coefficients $\ep_n$ are related to $p_n$ via the EOS.

In particular, the coefficients in front of $r^2$ are given by
\bea
0 &=& \mu_2 - 8 \frac{dF}{d\Ph} \mu_2 \Ph_2 - \frac{4\pi}{3} \ep_0 - \frac{1}{12} U ,
\nonumber \\
0 &=& - \frac{1}{2} \mu_2 + \nu_2  - 8 \frac{dF}{d\Ph} \nu_2 \Ph_2  - 2\pi p_0 + \frac{1}{8} U ,
\nonumber \\
0 &=&  - 8 \frac{dF}{d\Ph} \mu_2 \nu_2 + \Ph_2 - \frac{1}{12} \frac{dU}{d\Ph} ,
\nonumber \\
0 &=& (\ep_0 + p_0) \nu_2 + p_2 ,
\label{2ndrec}
\eea
where $dF/d\Ph, \, U$ and $dU/d\Ph$ are understood as their values at $\Ph = \Ph_0$. For a given pair of $p_0$ and $\Ph_0$, $\mu_2, \, \nu_2, \, p_2$ and $\Ph_2$ can be solved. However, noticing that the first three equations in \rf{2ndrec} are nonlinear, real solutions might not exist. In fact, eliminating two variables among $\{\mu_2, \, \nu_2, \, \Ph_2 \}$ leads to a quartic equation for the third one. For example, by eliminating $\nu_2$ and $\Ph_2$, we find
\bea
\mu_2^4 + a_3 \, \mu_2^3 + a_1 \, \mu_2 + a_0 = 0 ,
\label{quarticeq}
\eea  
with 
\bea
a_0 &=& \frac{1}{4608} \left( 16\pi \ep_0 + U \right)^2 \left( \frac{dF}{d\Ph} \right)^{-2} ,
\nonumber \\
a_1 &=& \frac{1}{1152 } \left( 16\pi \ep_0 + U \right) \left(2 \frac{dF}{d\Ph} \frac{dU}{d\Ph} - 3 \right) \left( \frac{dF}{d\Ph} \right)^{-2} ,
\nonumber \\
a_3 &=& 4\pi p_0 - \frac{1}{4} U .
\eea
Discussing the analytical conditions on $p_0$ and $\Ph_0$ for Eq.~\rf{quarticeq} to have real solutions would be too involved. Equation \rf{quarticeq} is practically solved numerically. We just point out that these conditions diminish the parameter space for spontaneous scalarization, causing abrupt stops in $\Ph_0$ as $p_0$ varies as shown in Fig.~\ref{fig_nonlin_Mbphc}.  

We would also like to point out that the center behavior of the linearized scalar field can be obtained from Eqs.~\rf{2ndrec} as a byproduct. Dropping all the $\Ph$-related quantities in the first and the second equations in \rf{2ndrec} recovers $\mu_2$ and $\nu_2$ in GR, namely
\bea
\mu_2 = \frac{4\pi}{3} \ep_0, \quad \nu_2 = \frac{2\pi}{3} \left( \ep_0 + 3p_0 \right) .
\eea 
Then, the third equation in \rf{2ndrec} gives $\Ph_2$ in the linearized case
\bea
\Ph_2 = \left[ \frac{128\pi^2 \xi }{9} \ep_0 \left( \ep_0 + 3p_0 \right) + \frac{1}{6\lambdabar_\Ph^2} \right] \Ph_0,
\label{lincph}
\eea
where $dF/d\Ph = 2\xi \Ph_0$ and $dU/d\Ph = 2\Ph_0/\lambdabar_\Ph^2$ have been used. As the recurrence relation is linear, the linearized scalar field inside the star is fixed up to a scale factor for a given NS configuration in GR.

\section{Asymptotic expansion of the spherical solution}
\label{app2}
It is illustrative to first study the asymptotic expansion of the linearized scalar field by only considering the third equation in Eqs.~\rf{odes} with $\mu$ and $\nu$ given by the Schwarzschild metric,
\bea
\mu = -\nu = - \frac{1}{2}\ln{ \left( 1 - \frac{2M}{r} \right)}, 
\label{grmunu}
\eea
where $M$ is the Arnowitt-Deser-Misner mass of the star. The ansatz
\bea
\Ph = e^{\pm r/\lambdabar_\Ph} \Ph_\pm  = e^{\pm r/\lambdabar_\Ph} \, r^{s_\pm} \sum_{n=1}^{\infty} \left( \Ph_{\pm} \right)_{n}\frac{1}{r^n}  ,
\label{linphasy}
\eea
is to be used to find series solutions for $\Ph$, where the constants $s_\pm$ and the expansion coefficients $\left( \Ph_{\pm} \right)_{n}$ are expected to be determined once the ansatz is substituted into the linearized scalar equation together with the series expansions of $\mu$ and $\nu$ in Eq.~\rf{grmunu}. Indeed, the leading-order expansion of the linearized scalar equation provides
\bea
\left( s_\pm \mp \frac{M}{\lambdabar_\Ph} \right) (\Ph_\pm)_1 = 0, 
\eea 
implying $s_\pm = \pm M/\lambdabar_\Ph$, while higher-order expansion terms give equations generating $(\Ph_\pm)_n$ from $(\Ph_\pm)_1$:
\bea
(\Ph_\pm)_2 &=& \frac{M}{2} \left( 1 - 3s_\pm \right) (\Ph_\pm)_1,
\nonumber \\
(\Ph_\pm)_3 &=& \frac{M^2}{8} \frac{9s_\pm^3 - 16 s_\pm^2 + 5s_\pm - 2}{ s_\pm} (\Ph_\pm)_1, 
\nonumber \\
&\cdots & \, . 
\label{linphasyreceq}
\eea

A general solution for $\Ph$ is a linear combination of $e^{ r/\lambdabar_\Ph} \Ph_+$ and $e^{- r/\lambdabar_\Ph} \Ph_-$. But physical solutions require $\Ph$ to be suppressed at infinity, meaning that $\Ph_+$ has to be zero. This leaves us one free coefficient $(\Ph_-)_1$ in the series solution for $\Ph$ as $(\Ph_+)_1$ must be set to zero. Since the linearized scalar equation is homogeneous, the free coefficient $(\Ph_-)_1$ is merely a scale factor apart from which the exterior solution for $\Ph$ is completely fixed for a star with a given value of mass $M$. A suitable value of the NS mass at which the exterior solution of $\Ph$ connects smoothly to the interior solution of $\Ph$ marks a possible onset of spontaneous scalarization. 

Now the series expansion in Eq.~\rf{linphasy} is ready to be generalized to the solution for the full set of ODEs in Eqs.~\rf{odes}. The nonlinearity inevitably leads to terms containing $e^{\pm 2 r/\lambdabar_\Ph}, \, e^{\pm 3 r/\lambdabar_\Ph}, \, e^{\pm 4 r/\lambdabar_\Ph}, \, \cdots$, so the physical solution can be formally written as
\bea
\mu &=& \mu^{(0)} + e^{-r/\lambdabar_\Ph} r^{s_-} \mu^{(1)} + e^{-2r/\lambdabar_\Ph} r^{2s_-} \mu^{(2)} + \cdots \, ,
\nonumber \\
\nu &=& \nu^{(0)} + e^{-r/\lambdabar_\Ph} r^{s_-} \nu^{(1)} + e^{-2r/\lambdabar_\Ph} r^{2s_-} \nu^{(2)} + \cdots \, ,
\nonumber \\
\Ph &=& e^{-r/\lambdabar_\Ph} r^{s_-} \Ph^{(1)} + e^{-2r/\lambdabar_\Ph} r^{2s_-} \Ph^{(2)} + \cdots \, ,
\label{nonlinasy}
\eea   
where $\mu^{(i)}, \, \nu^{(i)}$ and $\Ph^{(i)}$ have Taylor expansions in terms of $1/r$. When the ansatz \rf{nonlinasy} is substituted into Eqs.~\rf{odes}, terms containing the same exponential factor are collected to generate equations by the orders of $e^{- r/\lambdabar_\Ph}$. Starting with the zeroth order of $e^{- r/\lambdabar_\Ph}$, we acquire the equations for $\mu^{(0)}$ and $\nu^{(0)}$ that are identical to the GR equations for $\mu$ and $\nu$ while the scalar equation trivially vanishes. Next, at the first order of $e^{- r/\lambdabar_\Ph}$, the equations for $\mu^{(1)}, \, \nu^{(1)}$ and $\Ph^{(1)}$ come out. The equation for $\Ph^{(1)}$ is identical to the linearized scalar equation. Setting up their Taylor expansions in terms of $1/r$ and substituting into their equations, we find recurrence equations that fix the expansion coefficients of $\mu^{(1)}$ and $\nu^{(1)}$ to zero. Similar procedures apply to orders at $e^{-2r/\lambdabar_\Ph}, \, e^{-3r/\lambdabar_\Ph}, \, \cdots $, so the Taylor expansions for higher-order $\mu^{(i)}, \nu^{(i)}$ and $\Ph^{(i)}$ can be found.

We present the first few terms of the result:
\bw
\bea
g_{tt} &=& -1 + \frac{2M}{r}  +  \left[ \frac{(\Ph_-)_1}{M} \right]^2 e^{-2r/\lambdabar_\Ph} r^{2s_-} \left( \frac{1}{4} \frac{M^2}{r^2} + \cdots \right) + \cdots \, ,
\nonumber \\
g_{rr} &=& 1 + \frac{2M}{r} + \frac{4M^2}{r^2} + \frac{8M^3}{r^3} + \cdots + \left[ \frac{(\Ph_-)_1}{M} \right]^2 e^{-2r/\lambdabar_\Ph} r^{2s_-} \left( \frac{s_-}{2} \frac{M}{r} - \frac{3s_-^2-4s_-+1}{2} \frac{M^2}{r^2} + \cdots \right) + \cdots \, ,
\nonumber \\
\Ph &=& \frac{(\Ph_-)_1}{M} e^{-r/\lambdabar_\Ph} r^{s_-} \left( \frac{M}{r} + \frac{1-3s_-}{2} \frac{M^2}{r^2} + \frac{9s_-^3-16s_-^2+5s_--2}{8s_-} \frac{M^3}{r^3} + \cdots \right) + \left[\frac{(\Ph_-)_1}{M}\right]^3 e^{-3r/\lambdabar_\Ph} r^{3s_-} \left( \frac{s_-}{8} \frac{M^2}{r^2} + \cdots \right) + \cdots \, .
\label{nonlinasysol}
\eea
\ew
Here $M$ and $(\Ph_-)_1$ are the only two free parameters describing the asymptotic behavior of the solution. The constant $s_- = -M/\lambdabar_\Ph$ inherits the notation from the linearized case. The first few terms shown in Eqs.~\rf{nonlinasysol} provide a decent approximation for the solution when $r \gg M$ and $r \gg \lambdabar_\Ph$. By assuming that $r$ is much greater than the larger one of $M$ and $\lambdabar_\Ph$, the asymptotic behaviors of $g_{tt}$ and $g_{rr}$ are basically the same as those in GR since the modifications from the scalar field are proportional to the exponential factors and therefore are dropped. In astrophysical applications, the modifications from the scalar field at the far-field region however might not be dropped when $\lambdabar_\Ph \gg M$ so the suppression from the exponential factors is inefficient in the region $M \ll r \ll \lambdabar_\Ph$.
Noticing that the leading term with the exponential factor $e^{-2r/\lambdabar_\Ph}$ in $g_{tt}$ roughly has the same magnitude as the leading term of $\Ph^2$, a criterion to determine whether the scalar field affects the far-field metric significantly is checking if $\Ph$ decreases fast enough so that in the problematic region $M \ll r \ll \lambdabar_\Ph$ its value is much less than $\sqrt{M/r}$ or not. 

At the end of the discussion on the asymptotic behavior of the scalarized solution to Eqs.~\rf{odes}, we especially would like to point out that by setting $U = 0$ in Eqs.~\rf{odes} and employing Taylor expansions for $\mu, \, \nu$ and $\Ph$ in terms of $1/r$, their asymptotic expansions in the massless-scalar theory can be obtained. They turn out to be
\bea
g_{tt} &=& -1 + \frac{2M}{r}  - \frac{Q_s^2}{6} \frac{M^3}{r^3} + \cdots \, ,
\nonumber \\
g_{rr} &=& 1 + \frac{2M}{r} + \left( 4 - Q_s^2 \right) \frac{M^2}{r^2} + \left( 8 - \frac{5Q_s^2}{2} \right) \frac{M^3}{r^3} + \cdots \, ,
\nonumber \\
\Ph &=& Q_s \left( \frac{M}{r} + \frac{M^2}{r^2} + \left( \frac{4}{3} - \frac{Q_s^2}{12} \right) \frac{M^3}{r^3} + \cdots \right) \, ,
\label{linasysol}
\eea   
where $M$ and $Q_s$ are the two free parameters. It is surprising to see that Eqs.~\rf{linasysol} are not the limits of Eqs.~\rf{nonlinasysol} at $\lambdabar_\Ph \rightarrow \infty$. In fact, Eqs.~\rf{nonlinasysol} do not have limits at $\lambdabar_\Ph \rightarrow \infty$, because these infinite series contain increasing powers of $s_-$ in the denominators that unfortunately give rise to infinities unable to be regularized by redefining the free parameter $(\Ph_-)_1$ as $\lambdabar_\Ph \rightarrow \infty$. The discontinuity between the solutions of the massive scalar and of the massless scalar resembles the famous van Dam-Veltman-Zakharov discontinuity regarding the mass of graviton \cite{vanDam:1970vg, Zakharov:1970cc}. Apart from that, we have to admit that a plausible physical explanation for this puzzle as well as its rigorous mathematical resolution, is currently missing. 


\section{Equations for even-parity perturbations with $l \ge 2$}
\label{app3}
In this appendix, we present the equations for the radial functions $H_0, \, H_1, \, H_2, \, K, \, W_1, \, V$ and $X$ with the time dependence assumed to be $e^{i\si t}$ for the angular momentum number $l\ge 2$. When $l=0$ or $l=1$, certain components of the Regge-Wheeler gauge become trivial, suggesting that the freedom of choosing coordinates is incompletely utilized. Further coordinate transformations can eliminate one (for $l=1$) or two (for $l=0$) functions among $H_0, \, H_1, \, H_2, \, K, \, W_1$ and $V$, leading to different sets of equations from Eqs.~\rf{evenodes} displayed below (e.g., see Refs.~\cite{Chandrasekhar:1964zz, 1970ApJ...159..847C} for the GR case). Here we only focus on the case of $l \ge 2$ where the Regge-Wheeler gauge completely fixes the coordinates at the order of the perturbation. 

First, the Eulerian perturbation of the 4-velocity is found to be 
\bea
\de u^t &=& \frac{1}{2} e^{-\nu} H_0 Y_{lm} \, e^{i\si t} ,
\nonumber \\
\de u^r &=&  i\si e^{-\nu-2\mu}W_1 Y_{lm} \, e^{i\si t} ,
\nonumber \\
\de u^\th &=&  i\si e^{-\nu} V \frac{ \prt_\th Y_{lm}}{r^2} \, e^{i\si t} , 
\nonumber \\
\de u^\ph &=& i\si e^{-\nu} V \frac{ \prt_\ph Y_{lm} }{r^2 \sin^2\th} \, e^{i\si t} ,
\eea
by substituting Eqs.~\rf{evenper} into Eqs.~\rf{peru}. The covariant components 
\bea
\de u_t &=& \frac{1}{2} e^{\nu} H_0 Y_{lm} \, e^{i\si t} ,
\nonumber \\
\de u_r &=&  e^{-\nu} \left(H_1 + i\si W_1\right) Y_{lm} \, e^{i\si t} ,
\nonumber \\
\de u_\th &=&  i\si e^{-\nu} V \prt_\th Y_{lm} \, e^{i\si t} , 
\nonumber \\
\de u_\ph &=& i\si e^{-\nu} V \prt_\ph Y_{lm} \, e^{i\si t} ,
\eea
are then calculated using 
\bea
\de u_\al = g^{(0)}_{\al\be} \de u^\be + \de g_{\al\be} u^{(0)\be},
\eea
where $g^{(0)}_{\al\be} = {\rm diag}\Big\{ -e^{2\nu}, \, e^{2\mu}, \, r^2, \, r^2\sin^2\th \Big\}$ and $u^{(0)\be} = \left( e^{-\nu}, 0, 0, 0 \right)$ are the zeroth-order quantities.

Next, the Lagrangian perturbation of the particle number density is found to be
\bea
\frac{\De n}{n} = \left(\frac{1}{2} H_2 + K - \frac{l(l+1)}{r^2} V - \frac{e^{-\mu}}{r^2} W' \right) Y_{lm} e^{i\si t} ,
\label{evendeltan}
\eea  
where $W := -r^2 e^{-\mu} W_1$ is the fluid displacement function used by \citet{1967ApJ...149..591T} and the prime denotes the derivative with respect to $r$, by substituting Eqs.~\rf{evenper} into the second equation in Eqs.~\rf{perun}. Using Eqs.~\rf{perepp}, the Eulerian perturbations of the energy density and the pressure are then 
\bea
\de \ep &=& (\ep+p) \frac{\De n}{n} - e^{-2\mu} W_1 \ep' Y_{lm} e^{i\si t} ,
\nonumber \\
\de p &=& \ga p \frac{\De n}{n} - e^{-2\mu} W_1 p' Y_{lm} e^{i\si t} .
\eea 

Finally, the perturbative field equations can be calculated using the expressions of $\de g_{\al\be}, \, \de u_\al, \de \ep, \ \de p$ and $\de \Ph$. For the modified Einstein equations, the perturbation
\bea
\de E_\mn := \de G_\mn - 8\pi (\de T_m)_\mn - (\de T_{\Ph})_\mn - (\de T_{\rm GB})_\mn ,
\label{perefe}
\eea 
can be grouped into 4 scalars, 2 vectors, and 1 traceless tensor, providing 7 equations. Together with the perturbation of the scalar field equation, there are 8 equations in total for the 7 variables $H_0, \, H_1, \, H_2, \, K, \, W_1, \, V$ and $X$, indicating that one equation depends on the others. Please note that $W_0$ never shows up in the field equations. This is expected as $\xi_t$ is unnecessary in describing the motion of the fluid. 

The following is a set of 7 equations from which $H_0, \, H_1, \, H_2, \, K, \, W_1, \, V$ and $X$ can be solved:
\bw
\bea
0 &=& H_1 \left[ 1 - 2 r \mu' + \frac{(l+2)(l-1)}{2} e^{2\mu} + 8\pi r^2 e^{2\mu} \ep + \frac{1}{2} r^2 e^{2\mu} U + \frac{1}{2} r^2 \Ph^{\prime\, 2} + 4 \left(1-e^{-2 \mu}\right) F'' - \frac{2l (l+1)}{r} F' - 4  \left(1-3e^{-2 \mu}\right) \mu' F' \right]
 \nonumber \\
 && +  i \si H_2 \left[r e^{2 \mu} + 2 \left(e^{2 \mu}-3\right) F' \right] - i \si K'( r^2 e^{2 \mu} - 4 r F' )  + i \si K \left(r \nu'-1\right) \left(r e^{2 \mu}-4 F'\right) + 8\pi i \si W_1 r^2 e^{2 \mu} (\ep + p ) 
 \nonumber \\
 && + 4 i \si X'(1-e^{2 \mu}) \frac{dF}{d\Ph} + i \si X \left[ 4 (1-e^{2 \mu}) \Ph' \frac{d^2F}{d\Ph^2} - 4(1 - e^{2 \mu}) \nu' \frac{dF}{d\Ph} - r^2 e^{2 \mu} \Ph' \right]  ,
\nonumber \\
0 &=& H_1' \left( re^{2 \mu} - 4F' \right) + H_1 \left[ r e^{2 \mu} (\nu'- \mu') + 4 \left( 3 F' \mu' - F' \nu'+ \frac{1}{r} F' - F'' \right) \right]  - i \si H_2 e^{2 \mu} \left(r e^{2 \mu} - 4 F'\right) 
\nonumber \\
&&  - i \si K r e^{2 \mu} \left( e^{2 \mu} - 4 F'' + 4 F' \mu' \right) + 16\pi  i \si V r e^{4 \mu} ( \ep + p )  - 8i \si X e^{2 \mu} \mu' \frac{dF}{d\Ph}
\nonumber \\
0 &=& H_0' \left[ r + 2 \left(1 - 3 e^{-2 \mu}\right) F' \right]
- \frac{l(l+1)}{2} H_0  \frac{1 }{r} \left(r e^{2 \mu}-4 F'\right)
+ H_2  \left[ e^{2 \mu} + 4\pi r^2 e^{2 \mu} (2-\ga) p - \frac{1}{2} r^2 e^{2 \mu} U - 12 e^{-2 \mu} \nu' F' \right]
\nonumber \\
&& - K' r \left( 1 + r \nu' - 12 e^{-2 \mu} \nu' F' \right)
- K  \left[ \si^2 r^2 e^{2(\mu-\nu)} - \frac{(l+2)(l-1)}{2} e^{2 \mu} + 4\pi r^2 e^{2 \mu} \ga p - 4\si^2 r e^{-2 \nu} F' + 2 (l+2)(l-1) \nu' F' \right]
\nonumber \\
&& - 2 i \si H_1 e^{-2 \nu} \left[r+ 2 \left(1-3e^{-2\mu}\right) F' \right] 
- 4\pi W_1' r^2  \ga p  
- 4\pi W_1 r \left(  r p' + 2 \ga p - r \mu' \ga p \right)
+ 4\pi l (l+1) V e^{2 \mu} \ga p 
\nonumber \\
&& + X' \left[ r^2 \Ph' + 4 \left(1-3 e^{-2 \mu}\right) \nu' \frac{dF}{d\Ph} \right]
+ X \left[ 4 \si^2 e^{-2 \nu} \left(1 - e^{2 \mu}\right) \frac{dF}{d\Ph} - \frac{4l (l+1)}{r}  \nu'  \frac{dF}{d\Ph} - 4 \left(1-3e^{-2 \mu} \right) \nu' \Ph' \frac{d^2F}{d\Ph^2} - \frac{1}{2} \frac{dU}{d\Ph} r^2 e^{2 \mu} \right]
\nonumber \\
0 &=&  H_0' \left(r e^{2 \mu}-4 F'\right) + H_0 \frac{1}{r} \left(r \nu'-1\right) \left(r e^{2 \mu}-4 F'\right)  - i \si H_1 e^{-2 \nu} \left(r e^{2 \mu}-4 F'\right) + H_2 \left[ (1+r \nu') e^{2 \mu} - 12 \nu'  F' \right] 
\nonumber \\
 && - K' \left(re^{2 \mu}-4r \nu'F' \right) + 8 X' \nu' \frac{dF}{d\Ph} + 2 X \left( 4 \nu' \Ph' \frac{d^2F}{d\Ph^2} - 4 \frac{\nu'}{r} \frac{dF}{d\Ph} - r e^{2 \mu} \Ph' \right)
\nonumber \\
0 &=& H_0 \left(e^{2 \mu} - 4 F'' + 4 \mu' F' \right) - H_2 \left(e^{2 \mu}-4 \nu' F' \right) + 8 X \left(\nu''+\nu^{\prime\,2}-\mu' \nu'\right) \frac{dF}{d\Ph} ,
\nonumber \\
0 &=& H_0 r^2 (\ep+p) + H_2 r^2 \ga p +2 K r^2 \ga p + 2 W_1' r^2 e^{- 2 \mu} \ga p + 2W_1 r e^{- 2 \mu}\left( r p' + 2 \ga p - r \mu' \ga p \right) + 2V \left[ \si^2 r^2 e^{-2 \nu} (\ep+p) - l (l+1) \ga p \right] ,
\nonumber \\
0 &=& 2 H_0'' \left(1-e^{-2 \mu}\right) \frac{dF}{d\Ph}  
+ H_0'  \left[ 4 \left(1-e^{-2 \mu}\right) \nu' \frac{dF}{d\Ph} - 2 \left(1-3e^{-2 \mu}\right) \mu' \frac{dF}{d\Ph} - \frac{1}{2} r^2 \Ph^{\prime}  \right] 
- 2l (l+1) H_0 \frac{\mu'}{r} \frac{dF}{d\Ph}
\nonumber \\
&& - 4i\si H_1'  e^{-2 \nu}  \left(1-e^{-2 \mu}\right) \frac{dF}{d\Ph} 
+ i \si H_1  e^{-2 \nu} \left(r^2 \Ph^{\prime} + 4 \left(1-3e^{-2 \mu}\right) \mu' \frac{dF}{d\Ph} \right) 
+ H_2' \left[2 \left(1-3e^{-2 \mu}\right) \nu' \frac{dF}{d\Ph} - \frac{1}{2} r^2 \Ph^{\prime}\right]
\nonumber \\
&& + H_2 \left[ 4 \left(1-2e^{-2\mu}\right) \nu'' + 4 \left(1-2e^{-2\mu}\right) \nu^{\prime\,2} - 4 \left(1-6e^{-2\mu}\right) \mu' \nu' - \frac{2l (l+1)}{r} \nu' - 2 \si^2 \left(e^{2 \mu}-1\right) \right] \frac{dF}{d\Ph} 
\nonumber \\ 
&& + H_2  \left[ r \left(r \mu'-r \nu'-2\right) \Ph^{\prime}  - r^2 \Ph'' \right] + 4 K'' r e^{-2 \mu} \nu' \frac{dF}{d\Ph}
+ K'  \left[r^2 \Ph^{\prime} + 4  e^{-2 \mu} \left(r \nu''+r \nu^{\prime\,2} + 2 \nu' - 3 r \mu' \nu' \right) \frac{dF}{d\Ph} \right]
\nonumber \\
&& -2 K \left[ 2 \si^2 r e^{-2 \nu} \mu' + (l+2)(l-1) \left(\nu''+\nu^{\prime\,2} - \mu' \nu'\right)\right] \frac{dF}{d\Ph}  + X'' r^2 
+ X' r  \left( r \nu'-r \mu'+2 \right) 
\nonumber \\
&& + X \left[ \si^2  r^2 e^{2(\mu-\nu)} - l(l+1) e^{2 \mu}  + 4 \left(1-3e^{-2\mu}\right) \mu' \nu' \frac{d^2F}{d\Ph^2} - 4 \left(1-e^{-2\mu}\right) ( \nu'' + \nu^{\prime\,2} ) \frac{d^2F}{d\Ph^2}  - \frac{1}{2} r^2 e^{2 \mu} \frac{d^2U}{d\Ph^2} \right] .
\label{evenodes}
\eea
\ew
The first 5 equations are extracted from $\de E_{tr}, \, \de E_{t\th}, \, \de E_{rr}, \, \de E_{r\th}$ and $\de E_{\th\ph}$. The penultimate equation is extracted from the $\th$-component of the conservation equation \rf{coneq}, rather than from $\de E_{tt}$ or $\de E_{\th\th}+\de E_{\ph\ph}/\sin^2\th$ as they are even more lengthy. The last equation is from the perturbative scalar field equation. We notice that all the equations are independent of the angular momentum number $m$ while some of them depend on $l$. The condition $l\ge 2$ should be addressed again here, because $l\ge 1$ has to be assumed to extract the equations from the vector components $\de E_{t\th}$ and $\de E_{r\th}$, and $l\ge 2$ has to be assumed to extract the equation from the tensor component $\de E_{\th\ph}$.

Among the 7 perturbative functions, $H_1, \, H_2$ and $V$ can be solved in terms of other functions using the first, the fifth, and the sixth equations in Eqs.~\rf{evenodes}. So the system can be reduced to 4 ODEs for $H_0, \, K, \, W_1$ and $X$, with the highest derivatives involved being $H_0', \, K', \, W_1''$ and $X''$.
The boundary conditions imposed to get physical solutions are analogous to those in GR. At the center of the star, the functions have Taylor expansions taking the following form  
\bea
H_0 &=& H_{00} r^{l} + H_{02}r^{l+2} + \cdots \, ,
\nonumber \\
H_1 &=& H_{10} r^{l+1} + H_{12} r^{l+3} \cdots \, ,
\nonumber \\
H_2 &=& H_{20} r^{l} + H_{22} r^{l+2} \cdots \, ,
\nonumber \\
K &=& K_{0} r^{l} + K_{2} r^{l+2} \cdots \, ,
\nonumber \\
W_1 &=& W_{10} r^{l-1} + W_{12} r^{l+1} \cdots \, ,
\nonumber \\
V &=& V_{0} r^{l} + V_{2} r^{l+2} \cdots \, ,
\nonumber \\
X &=& X_{0} r^{l} + X_{2} r^{l+2} \cdots \, ,
\label{evencbc}
\eea
where only three of the coefficients are free. For example, using $H_{00}, \, W_{10}$ and $X_0$ as the free coefficients, we can find  
\bea
H_{10} &=& \frac{2i\si}{l+1} \left[  H_{00} \left(1- 8 \Ph_2 \frac{dF}{d\Ph}\right) + 16 X_0 \nu_2 \frac{dF}{d\Ph}   \right]
\nonumber \\
&&  + \frac{16i\si}{l+1} \left[ \frac{ X_0 \mu_2 \frac{dF}{d\Ph} }{1- 8 \Ph_2 \frac{dF}{d\Ph}} - \frac{ \pi W_{10} \left( \ep_0 + p_0 \right)}{ l \left(1- 8 \Ph_2 \frac{dF}{d\Ph}\right) } \right] \, ,
\nonumber \\
H_{20} &=& K_0 = H_{00} \left(1 - 8 \Ph_2 \frac{dF}{d\Ph} \right) + 16 X_0 \nu_2 \frac{dF}{d\Ph} \, ,
\nonumber \\
V_0 &=& \frac{ W_{10} }{l} ,
\label{evencbc0}
\eea
where $\mu_2, \, \nu_2$ and $\Ph_2$ are solutions from Eqs.~\rf{2ndrec} and $dF/d\Ph$ is understood to take its value at the center of the star, as well as higher-order recurrence equations to generate the higher-order coefficients. At infinity, the solutions for the perturbative functions feature the factors $e^{\pm i\si r}$ while those with the factor $e^{ i\si r}$ need to be discarded to fulfill the boundary condition of no incoming waves taking into consideration of the assumed time-dependence $e^{i\si t}$. An extra boundary condition at infinity compared to GR comes from the fact that the factors $e^{\pm r/\lambdabar_\Ph}, \, e^{\pm 2r/\lambdabar_\Ph}, \, \cdots$, appear in the solutions for the perturbative functions, where $\lambdabar_\Ph$ is the reduced Compton wavelength of the scalar field, so those with the factors $e^{ r/\lambdabar_\Ph}, \, e^{ 2r/\lambdabar_\Ph}, \, \cdots $, need to be discarded for the perturbation to be consistent with the spherical solution. Last but not least, at the surface of the star every physical solution has a vanishing Lagrangian perturbation of the particle number density, leaving a condition on the combination of the functions $H_2, \, K, \, V$ and $W_1$ as shown in Eq.~\rf{evendeltan}.  

For the numerical approach that integrates from the center of the star to solve Eqs.~\rf{evenodes}, the use of the Taylor expansions \rf{evencbc} automatically meets the boundary condition at the center. To satisfy the boundary conditions at the surface of the star and at infinity, the free parameters $H_{00}, \, W_{10}, \, X_{0}$ as well as the complex frequency $\si$ need to be adjusted to suitable values. Noticing that Eqs.~\rf{evenodes} are homogeneous for the set of the perturbative functions, it is the ratios between $H_{00}, \, W_{10}, \, X_{0}$ that matter. After taking $H_{00} = 1$, the shooting strategy can be implemented for finding suitable values of $W_{10}$ and $X_0$, aiming at vanishing $\De n$ at the surface of the star and the divergent part in $\de \Ph$ at infinity caused by the factors of $e^{ r/\lambdabar_\Ph}$. The suitable values of $\si$, namely the quasinormal frequencies, are to be found by shooting for eliminating the $e^{i\si r}$ part of the solutions in principle, but the process of trial and error is much elaborate due to the divergence in the solutions at infinity caused by the imaginary part of $\si$ \cite{Lindblom:1983ps, Detweiler:1985zz, Chandrasekhar:1991fi, 1993CQGra..10..735A}.

For the static case studied in this work, the solutions are much simplified by setting $\si = 0$. The first equation in Eqs.~\rf{evenodes} gives $H_1 = 0$, and remarkably, $W_1$ and $W_1'$ can be eliminated altogether with $V$ using the sixth equation, leaving us two first-order ODEs for $H_0$ and $K$, and a second-order ODE for $X$. We only need to find suitable values of one parameter, namely $X_0$, using the shooting strategy. In fact, following \citet{Pani:2014jra}, due to the linearity of Eqs.~\rf{evenodes} in terms of the perturbative functions, instead of applying the shooting strategy two independent solutions with $H_{00} = 1, \, X_0 = 0$ and $H_{00} = 1, \, X_0 = 1$ are computed first, and the physical solution with $X$ vanishing at infinity can be conveniently constructed as a linear combination of the two computed solutions.

\section{Calculating the tidal deformability using the formula of Hinderer }
\label{app4}
Using the asymptotic expressions \rf{nonlinasysol} for the spherical solution we can obtain the following asymptotic expansion for $H_0$ when $\si = 0$ and $l=2$: 
\bw
\bea
H_0 &\rightarrow& a_P \left( -\frac{3r^2}{M^2} + \frac{6r}{M} \right) 
+ a_Q \left( -\frac{8}{5} \frac{M^3}{r^3} - \frac{24}{5} \frac{M^4}{r^4} + \cdots \right) 
 + \left[ \frac{\left( \Ph_- \right)_1}{M} \right]^2 e^{-2r/\lambdabar_\Ph} r^{2s_-} \left[ a_P \left( -\frac{s_-}{4} \frac{r}{M} + \cdots \right) + a_Q \left( \frac{s_-}{5} \frac{M^4}{r^4} + \cdots \right) \right] + \cdots ,
 \label{asyh02}
\eea
\ew
where $\left(\Ph_-\right)_1$ is the coefficient in the spherical solution for $\Ph$ and $s_- = - M/\lambdabar_\Ph$. To match the leading terms as shown in Eq.~\rf{asyh0tidal}, the coefficients $a_P$ and $a_Q$ are
\bea
a_P = \frac{M^2}{3} \frac{{\mathcal E}_{2m}}{A_m}, \quad a_Q = -\frac{15}{8M^3} \frac{Q_{2m}}{A_m} ,
\eea
so the tidal deformability can be calculated using $a_P$ and $a_Q$ as
\bea
\la = \frac{8}{45} \frac{a_Q}{a_P} M^5.
\label{tid2}
\eea 
Similarly to the spherical solution, the expansion for $H_0$ is simply the same as that in GR after dropping the terms involving the factors of $e^{-r/\lambdabar_\Ph}$. Since for static perturbations, $H_0$ in GR outside the star has the analytical solution
\bea
\left( H_0 \right)_{\rm GR} = a_P P^2_l\left( \frac{r}{M}-1 \right) + a_Q {\rm Re}\left[ Q^2_l\left( \frac{r}{M} - 1 \right) \right] ,
\label{grh0ext}
\eea
with $P^2_l$ and $Q^2_l$ being the associated Legendre functions, the coefficients $a_P$ and $a_Q$ can be calculated once the values of $H_0$ and its derivative are given at a sufficiently large radius $r_\infty$, namely
\bea
a_P &=& \frac{H_0' {\rm Re}\left[Q^2_l \right] - H_0 {\rm Re}\left[Q^2_l \right]' }{ \left( P^2_l \right)' {\rm Re}\left[Q^2_l \right] - P^2_l {\rm Re}\left[Q^2_l \right]'} \Bigg|_{r=r_\infty},
\nonumber \\
a_Q &=&  -\frac{H_0' P^2_l - H_0 \left( P^2_l \right)'}{ \left( P^2_l \right)' {\rm Re}\left[Q^2_l \right] - P^2_l {\rm Re}\left[Q^2_l \right]' }\Bigg|_{r=r_\infty} , 
\label{apaq}
\eea   
where the prime denotes the derivative with respect to $r$ and the argument of the associated Legendre functions is $r/M-1$. Taking $l=2$ and defining $y := r H_0'/H_0 \big|_{r=r_\infty}$, $C := M/r_\infty $, the tidal deformability given by substituting Eqs.~\rf{apaq} into Eq.~\rf{tid2} has exactly the same form as the formula of \citet{Hinderer:2007mb}
\bea
\la &=& \frac{16}{15} M^5 (1-2 C)^2 \left[ 2 C y - 2C + 2-y\right] 
\nonumber \\
&& \times \Big[ 8 C^5 (y+1)+ 4 C^4(3 y-2) + 4 C^3 (13-11 y)
\nonumber \\
&& \hskip 5pt + 6 C^2 (5 y-8)+ 2 C (6-3 y) 
\nonumber \\
&& \hskip 5pt + 3 (1-2 C)^2 ( 2 C y - 2C +2-y ) \ln \left(1-2C\right) \Big]^{-1} .
\label{tid3}
\eea

\begin{figure}
 \includegraphics[width=0.9\linewidth]{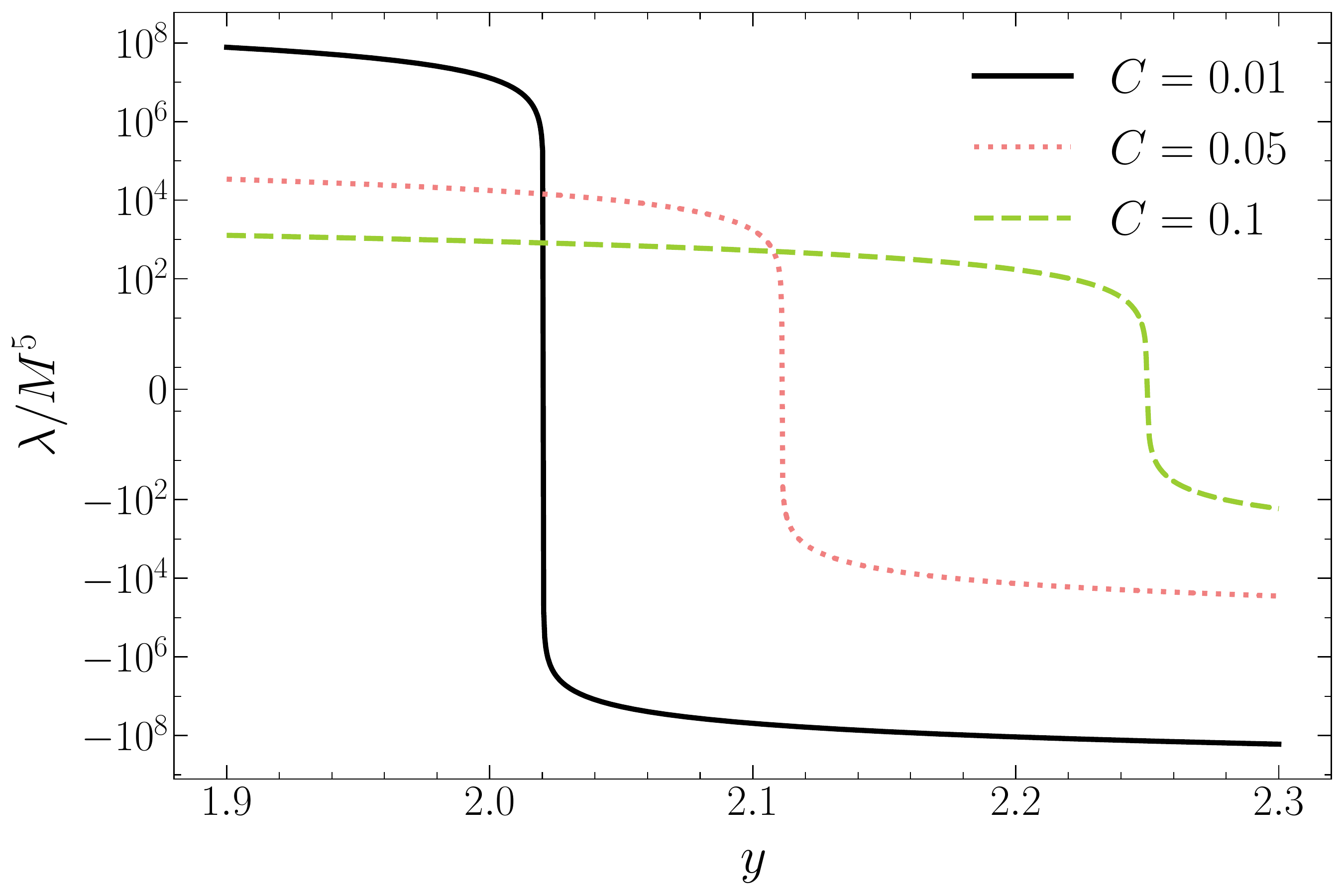}
 \caption{$\lambda/M^5$ versus $y$ for different values of $C$.   }
\label{fig_app4}
\end{figure} 

Now supposing $\lambdabar_\Ph \gg M$, we take $r_\infty \approx \lambdabar_\Ph$ for the exponential term in Eq.~\rf{asyh02} to drop. Then the issue of $C \approx M/\lambdabar_\Ph $ being too small is exposed in using the formula \rf{tid3}. Simply plotting $\lambda$ versus $y$ for several values of $C$ in Fig.~\ref{fig_app4} reveals the rather violent change of $\lambda$ when $y$ is near its asymptotic value $2$ and $C\lsim 0.01$. Remarkably, the error in $\la$ caused by an inaccurate $y$ with an error $\de y$ can be estimated by
\bea
\de \la \approx \left| \frac{\prt\la}{\prt y} \right| \de y \approx \frac{M^5}{15 C^5} \de y,
\eea  
where $y \approx 2$ has been used. The error in $y$ comes from the error in computing the solution for $H_0$, which can be estimated by the error of the integrator. For the efficiency of our code, the relative error of the integrator has been set to $10^{-7}$. Using $\de y = y\times 10^{-7} \approx 2 \times 10^{-7}$, the absolute error in $\la$ is, for instance, $\de \la \approx 0.04\times M^5$ when $ M/\lambdabar_\Ph = 0.05$ and $\de \la \approx 133 \times M^5$ when $M/\lambdabar_\Ph = 0.01$. The mass of NSs is at the order of kilometer in geometrized units. So when $\lambdabar_\Ph\sim 100\,{\rm km}$ ($a=0.1$), the absolute error for the dimensionless tidal deformability is as high as $10^2$, which is unacceptable if $\lambda/M^5$ itself is at the same order of magnitude or even less.
\\


\section{Equations for odd-parity perturbations}
\label{app5}
In this appendix, we present the equations for the radial functions $h_0, \, h_1$ and $h_2$ with the time dependence assumed to be $e^{i\si t}$. Here the angular momentum number $l$ takes positive integers as the odd-parity spherical harmonics start from $l=1$.

Substituting Eqs.~\rf{oddper} into Eq.~\rf{peru}, the Eulerian perturbation of the 4-velocity turns out to be
\bea
\de u^t &=& \de u^r = 0,
\nonumber \\
\de u^\th &=& - i\si e^{-\nu} h_2  \frac{\prt_\ph Y_{lm} }{r^2 \sin\th} \, e^{i\si t} , 
\nonumber \\
\de u^\ph &=& i\si e^{-\nu} h_2 \frac{\prt_\th Y_{lm} }{r^2 \sin\th} \, e^{i\si t}  .
\label{oddu}
\eea
The covariant components are
\bea
\de u_t &=& \de u_r = 0,
\nonumber \\
\de u_\th &=& - e^{-\nu}( h_0 + i\si h_2)  \frac{\prt_\ph Y_{lm} }{\sin\th} \, e^{i\si t} , 
\nonumber \\
\de u_\ph &=& e^{-\nu} (h_0 + i\si h_2) \sin\th \, \prt_\th Y_{lm}  \, e^{i\si t}  .
\eea
Direct calculation using the second equation in Eqs.~\rf{perun} verifies that the Lagrangian perturbation of the particle number density vanishes, and hence 
\bea
\de \ep = \de p = 0.
\eea

The perturbative Einstein equations \rf{perefe} provide 3 equations for $h_0, \, h_1$ and $h_2$ when $l \ge 2$ but only 2 equations when $l=1$, consistent with the fact that the Regge-Wheeler gauge is trivial for the case of odd-parity with $l=1$. Using the $\th$-component of the conservation equation instead of the tensor component $\de E_{\th\ph}$ that is invalid when $l=1$, we can write down the following three equations:
\bw
\bea
0 &=& h_0'' \left( r^2 e^{2 \mu} - 4 r F' \right) 
- h_0' \left[ r^2 e^{2 \mu} (\mu' + \nu') + 4 r F''- 4 \left(1+3 r \mu'+r \nu'\right) F' \right] 
\nonumber \\
&& - h_0 \left[  2 e^{2 \mu}( 1 - 2 r \mu' + r^2 \nu''+ r^2 \nu^{\prime\,2} - r^2  \mu' \nu' ) + (l+2)(l-1)e^{4 \mu} + 16\pi r^2 e^{4 \mu} \ep \right]
\nonumber \\
&& - h_0 \left[  r^2 e^{2 \mu} \Ph^{\prime\,2} + r^2 e^{4 \mu} U  +  4 (l+2)(l-1) e^{2 \mu} ( \mu' F'- F'') + 8 \left( \nu' + 3 \mu' - r \nu'' - r \nu^{\prime\,2} + 3 r \mu' \nu' \right) F' - 8 ( 1 + r  \nu') F'' \right]
\nonumber \\
&& - i \si h_1'  r \left(r e^{2 \mu}-4 F'\right) 
+ i \si h_1  \left[  r^2 e^{2 \mu} ( r\mu' + r\nu'- 2 ) + 4 \left(1-3 r \mu'-r \nu' \right)F' + 4r F'' \right]  
- 16 \pi i \si h_2 r^2 e^{4\mu} (\ep+p),
\nonumber \\
0 &=& i \si h_0' \left(r^2 e^{2 \mu}-4r F'\right) - 2 i \si  h_0  \left(r e^{2 \mu}-4 F'\right) 
\nonumber \\
&& + h_1 \left[ \si^2 r^2 e^{2 \mu} - (l+2)(l-1) e^{2 (\mu+\nu)} - 2 r e^{2 \nu}( \nu' - \mu' + r\nu'' + r \nu^{\prime\,2} - r  \mu' \nu' ) + 16\pi r^2 e^{2 (\mu+\nu)} p - r^2 e^{2 \nu} \Ph^{\prime\,2} - r^2 e^{2 (\mu+\nu)} U\right] 
\nonumber \\
&& + 4 h_1 \left[ - \si^2 r\Ph^{\prime}\frac{dF}{d\Ph} + (l+2)(l-1) e^{2\nu} \nu' \Ph^{\prime}\frac{dF}{d\Ph} + 2 r e^{2(\nu-\mu)}( \nu''+ \nu^{\prime\,2}  - 3 \mu' \nu' ) \Ph^{\prime}\frac{dF}{d\Ph}  + 2 r e^{2 (\nu-\mu)} \nu' F'' \right] ,
\nonumber \\
0 &=& \si^2 h_2 - i\si h_0 ,
\label{oddodes}
\eea
\ew
where the first and the second equations come from $\de E_{t\th}$ and $\de E_{r\th}$, and the third equation is from the $\th$-component of the conservation equation. They are valid for $l$ taking all positive integers only that one of the equations becomes dependent on the other two when $l = 1$. 

For $\si \ne 0$, a coordinate transformation can be performed at the order of the perturbation to eliminate one of the perturbative functions in the case of $l=1$. Choosing it to be $h_1 = 0$, then the second equation in Eqs.~\rf{oddodes} immediately gives $h_2 \propto r^2$, which does not contain the oscillating factors $e^{\pm i \si r}$, implying no odd-parity dipole gravitational wave in the scalar-tensor theory. When $l \ge 2$, the Regge-Wheeler gauge completely fixes the coordinates and the three independent equations in Eqs.~\rf{oddodes} are sufficient to solve the three variables $h_0, \, h_1$ and $h_2$. The boundary conditions for physical solutions include the finite behavior at the center of the star described by the expansions 
\bea
h_0 &=& h_{00} r^{l+1} + h_{02} r^{l+3} + \cdots ,
\nonumber \\
h_1 &=& h_{10} r^{l+2} + h_{12} r^{l+4} + \cdots ,
\label{oddcbc1}
\eea
as well as the requirement of no incoming waves at infinity. Note that there is only one free coefficient in the expansions \rf{oddcbc1}. Specifically speaking, $h_{00}$ and $h_{10}$ are related via
\bea
(l+2) h_{10} = i\si e^{-2\nu} \left( 1 - 8 \Ph_2 \frac{dF}{d\Ph} \right) h_{00} ,
\eea  
where $\Ph_2$ is solved from Eqs.~\rf{2ndrec}, and $\nu$ and $dF/d\Ph$ are understood to take their values at the center of the star, while the higher-order coefficients are generated from $h_{00}$ and $h_{10}$ by recurrence equations. Since Eqs.~\rf{oddodes} are homogeneous in the perturbative functions, the free coefficient in the expansions \rf{oddcbc1} is merely a scale factor. Therefore the only parameter that requires suitable values for physical solutions is the frequency $\si$, which should be selected to eliminate the incoming wave at infinity.

For $\si=0$, the case of $l=1$ is not much different from the case of $l \ge 2$, because the perturbative Einstein equations only provide two independent equations even when $l \ge 2$ as long as $\si = 0$. This is conveniently revealed by the fact that the third equation in Eqs.~\rf{oddodes} trivially vanishes. The problem is further simplified as the second equation in Eqs.~\rf{oddodes} gives $h_1 = 0$. In conclusion, the physical solution to the first equation in Eqs.~\rf{oddodes} turns out to describe the only nonvanishing off-diagonal metric component
\bea
g_{t\ph} = h_0 \sin\th \, \prt_\th Y_{l0},
\eea
of a slowly rotating star. In fact, writing the 4-velocity of a slowly rotating star as 
\bea
u^\mu  = \frac{1}{ \sqrt{-g_{tt} - 2 \Om g_{t\ph} - \Om^2 g_{\ph\ph}} } \left( 1, 0, 0 , \Om \right),
\eea
and expanding it up to the linear order of the angular velocity $\Om$, the perturbation of the 4-velocity is found to be
\bea
\de u^\al = \left( 0, 0, 0, e^{-\nu} \Om \right),
\eea 
which matches Eqs.~\rf{oddu} by taking $m=0$ and
\bea
i\si h_2 \frac{\prt_\th Y_{l0}}{r^2 \sin\th} \rightarrow \Om,
\label{slowrot}
\eea
as $\si \rightarrow 0$. To solve $h_0$, the angular velocity $\Om$ needs to be specified. But it cannot be specified arbitrarily due to the match \rf{slowrot}. In general it takes the form
\bea
\Om = \tilde \Om \frac{\prt_\th Y_{l0}}{\sin\th},
\label{angularvel}
\eea
where $\tilde \Om$ is a given function of $r$ and has the expansion
\bea
\tilde \Om = \tilde \Om_{0} r^{l-1} + \tilde \Om_{2} r^{l+1} + \tilde \Om_{4} r^{l+3} + \cdots ,
\eea
at the center of the star. The solution of $h_0$ then still has an expansion in the form of the first equation in Eqs.~\rf{oddcbc1} at the center of the star, now with the coefficients $\tilde \Om_0, \, \tilde \Om_2, \, \tilde \Om_4, \, \cdots$, entering into the recurrence equations for $h_{0n}$. For example, $h_{02}$ is related to $h_{00}$ and $\tilde \Om_0$ via
\bea
h_{02} &=& \frac{ \left[ 4\pi\ep_0 (l+1) + 2\pi p_0(l+3) + \left(l+\frac{1}{2}\right)(l-1) \mu_2  \right]h_{00} }{ (2 l+3) (1-8 \Ph_2 \frac{dF}{d\Ph})}  
\nonumber \\
&& - \frac{(l-1) \left[ 16 (l+1)\Ph_4 \frac{dF}{d\Ph} + 8(l+1) \Ph_2^2 \frac{d^2F}{d\Ph^2} - \frac{1}{8}U \right] h_{00} }{(2 l+3) (1-8 \Ph_2 \frac{dF}{d\Ph})}
\nonumber \\
&& + \frac{  8\pi(\ep_0+p_0) \tilde \Om_0 }{(2 l+3) (1-8 \Ph_2 \frac{dF}{d\Ph})} ,
\label{oddstaticcbc}
\eea
where $p_0, \, \ep_0, \, \mu_2, \, \Ph_2$ and $\Ph_4$ are the expansion coefficients defined in Eqs.~\rf{cexp}, and $\frac{dF}{d\Ph} , \, \frac{d^2F}{d\Ph^2}$ and $U$ are understood to take their values at the center of the star. Differently from the case of $\si \ne 0$, now there are two free parameters, namely $h_{00}$ and $\tilde \Om_0$, and their ratio determines the solution of $h_0$, which in general consists of a part whose leading asymptotic term is proportional to $r^{l+1}$ and a part whose leading asymptotic term is proportional to $1/r^l$. For $g_{t\ph}$ to vanish at infinity, a suitable value of $h_{00}/\tilde \Om_0$ is demanded so that the $r^{l+1}$ part does not show up.

\bibliography{refs}

\end{document}